\documentclass{aa}
\usepackage{amssymb,amsfonts,amsmath,times,pict2e}
\usepackage{natbib}
\usepackage{graphicx}
\usepackage{enumerate}
\usepackage{subfigure}
\usepackage{bm}
\usepackage{amsmath,amsbsy}
\usepackage{underscore}
\usepackage{float}

\usepackage{color}
\usepackage{ulem}
\definecolor{mygray}{gray}{0.6}
\definecolor{magenta}{rgb}{0.858, 0.188, 0.478}

\newcommand{\Rmnum}[1]{\expandafter\@slowromancap\romannumeral #1}

\usepackage{xspace}
\newcommand{\fg}[1]{Fig.~\ref{fig:#1}}
\newcommand{\Fg}[1]{Figure~\ref{fig:#1}}

\newcommand{\fgnum}[1]{\ref{fig:#1}}
\newcommand{\eq}[1]{Eq.~(\ref{eq:#1})\xspace}
\newcommand{\Eq}[1]{Equation~(\ref{eq:#1})\xspace}
\newcommand{\eqs}[2]{Eqs.\ (\ref{eq:#1}) and (\ref{eq:#2})}

\newcommand{\eqss}[3]{Eqs.\ (\ref{eq:#1}), (\ref{eq:#2}) and (\ref{eq:#3})}

\newcommand{\tb}[1]{Table~\ref{tab:#1}\xspace}
\newcommand{\se}[1]{Sect.~\ref{sec:#1}\xspace}
\newcommand{\Se}[1]{Section~\ref{sec:#1}\xspace}
\newcommand{\ses}[2]{Sects.\ \ref{sec:#1} and \ref{sec:#2}}
\newcommand{\ap}[1]{Appendix ~\ref{ap:#1}\xspace}

\newcommand{\ie}{i.e.}

\newcommand{\eg}{e.g.}

\newcommand{\AU}{ \  \rm AU}
\newcommand{\K}{ \  \rm K}
\newcommand{\Ms}{ \  \rm M_\odot }
\newcommand{\Msyr}{ \  \rm M_\odot \, yr^{-1} }

\newcommand{\Me}{ \ \rm M_\oplus}

\newcommand{\taus}{ \tau_{\rm s}}

\newcommand{\Omegak}{\Omega_{\rm K}}

\newcommand{\alphag}{\alpha_{\rm g}}
\newcommand{\alphat}{\alpha_{\rm t}}
\newcommand{\Mp}{M_{\rm p}}

\usepackage{CJKutf8}

\begin{document}

\title{Super-Earth masses sculpted by pebble isolation around  stars of different masses}
 
 \author{ Beibei Liu \inst{1},  Michiel Lambrechts \inst{1}, Anders Johansen \inst{1}, \and Fan, Liu \inst{1} }
 \authorrunning{B. Liu et al.}
\institute{
Lund Observatory, Department of Astronomy and Theoretical Physics, Lund University, Box 43, 22100 Lund, Sweden. \label{inst1} \\
\email{bbliu@astro.lu.se}
 }
\date{\today}

\abstract{
We develop a pebble-driven core accretion model to study the formation and evolution of  planets around stars in the stellar mass range of $0.08\Ms$ and $1\Ms$. 
By Monte Carlo sampling of the initial conditions, the growth and migration of a large number of individual  protoplanetary embryos are simulated in a population synthesis manner. We test two hypotheses  for the birth locations of embryos: at the water ice line or log-uniformly distributed over entire protoplanetary disks. Two types of disks with different turbulent viscous parameters $\alphat$ of $10^{-3}$ and $10^{-4}$ are also investigated, to shed light on the role of outwards migration of protoplanets. The forming planets are compared with the observed exoplanets in terms of masses, semimajor axes, metallicities and water contents. 
We find that gas giant planets are likely to form when the characteristic disk sizes are larger, the disk accretion rates are higher,  the disks are more metal rich and/or their stellar hosts are more massive. Our model shows that 1) the characteristic mass of super-Earth  is set by the pebble isolation mass. Super-Earth masses increase linearly with the mass of its stellar host, corresponding to one Earth mass around a late M-dwarf star and $20$ Earth masses around a solar-mass star. 2) The low-mass planets up to $20 \Me$ can form around stars with a wide range of metallicities, while massive gas giant planets are preferred to grow around metal rich stars. 3) Super-Earth planets that are mainly composed of silicates, with relatively low water fractions can form from protoplanetary embryos at the water ice line in weakly turbulent disks where outwards migration is suppressed. However, if the embryos are formed over a wide range of radial distances, the super-Earths would end up having a distinctive, bimodal composition in water mass. Altogether, our model succeeds in quantitatively reproducing several important observed properties of exoplanets and correlations with their stellar hosts. 
} 


\maketitle

\section{Introduction}
\label{sec:introduction}

The discovery and characterization of exoplanets is a rapidly evolving research field. Thanks to the increasing number of exoplanets with the masses ($M_{\rm p}$), radii ($R_{\rm p}$), semimajor axes ($a$), valuable constraints can be put for and test with the planet formation theory.

Until now, the most common type of known planets are super-Earths (sometimes also called mini-Neptunes, here defined as the planets with $1 \ R_{\oplus} \leq R_{\rm p} \leq  4 \ R_{\oplus}$  or $1 \Me \leq M_{\rm p} \lesssim 10-20 \Me$). Statistically, from the Kepler survey the occurrence rate of super-Earths is $\approx 30\%$ around solar-type stars  \citep{Fressin2013,Zhu2018} , and this rate is even higher around less massive M stars \citep{Mulders2015}. 
Planets within the multi-planet systems discovered by the Kepler satellite are found to have similar sizes and masses, than if the planets in the system were randomly assembled  from the observed population 
 (\citealt{Weiss2018,Millholland2017}; but also see \citealt{Zhu2019}). 
This intra-system uniformity suggests that a  given planetary system tends to have planets with the same characteristic mass.  

Recent studies further show some degree of the inter-system uniformity as well, \ie , planets from different systems around same type stars have  similar masses.  
\cite{Wu2018} suggested that the masses of Kepler planets are linearly scaled with their stellar masses (their Fig. 11), based on a detailed  analysis from the shift of photoevaporation valley to larger planets around more massive stars. Inferring the masses of Kepler planets from \cite{Chen2017}'s mass-radius relation,  \cite{Pascucci2018}  found the occurrence rate has a break at a planet-to-star mass ratio of $3 \times10^{-5}$ (their Fig. 1). This universal truncation is true for Kepler planets around stars below $1 \Ms$. Both findings support that planets in different systems tend to have a similar mass, and this characteristic mass linearly scales with the masses of their stellar hosts.
 Note, however, that the Kepler target stars are mostly in the range of early  M dwarfs and FGK stars. How well the above mentioned correlation can extend to the planetary systems around very low-mass stars is unknown.   
 
 Thanks to the improvement of ground-based photometric instruments, a growing number of planetary systems around very low-mass late M stars have been discovered in recent years.  
 For instance, seven Earth-sized planets with very compact and short period orbits were discovered around the ultra-cool TRAPPIST-$1$ star whose mass is only $0.08 \Ms$ \citep{Gillon2017,Luger2017}. 
 A $1.4 \ R_{\oplus}$, rocky super-Earth planet in the  habitable zone was detected to transit around $\rm LHS \ 1140$, a $0.15 \Ms$ star  \citep{Dittmann2017}.  The HARPS radial velocity survey also discovered a terrestrial planet $\rm Ross \  128b$ around a nearby $0.17\Ms$ dwarf \citep{Bonfils2018}, and a multiple Earth-mass planetary system around the late M-dwarf YZ  Ceti \citep{Astudillo-Defru2017}. 
An Earth-mass planet ($\rm OGLE$-$2016$-$\rm BLG$-$\rm 1195Lb$) in a $1$ AU orbit was also discovered around another ultracool dwarf star by microlensing surveys \citep{Shvartzvald2017}.  Recently, more and more such low-mass planets have been subsequently discovered around these late M-dwarf stars, such as Teegarden's  star \citep{Zechmeister2019},  GJ  $1265$ \citep{Luque2018}, GJ $1061$  \citep{Dreizler2019} and GJ  $357$ \citep{Luque2019}.
  The increasing observations raise the question of whether the terrestrial, earth-sized planets around  very low-mass stars ($\simeq 0.1 \Ms$) are analogous to but a scaled-down version of the Kepler systems with larger planets observed around more massive stars ($\sim 0.4-1.2 \Ms$)?

\begin{figure}[tbh]
 \includegraphics[scale=0.5, angle=0]{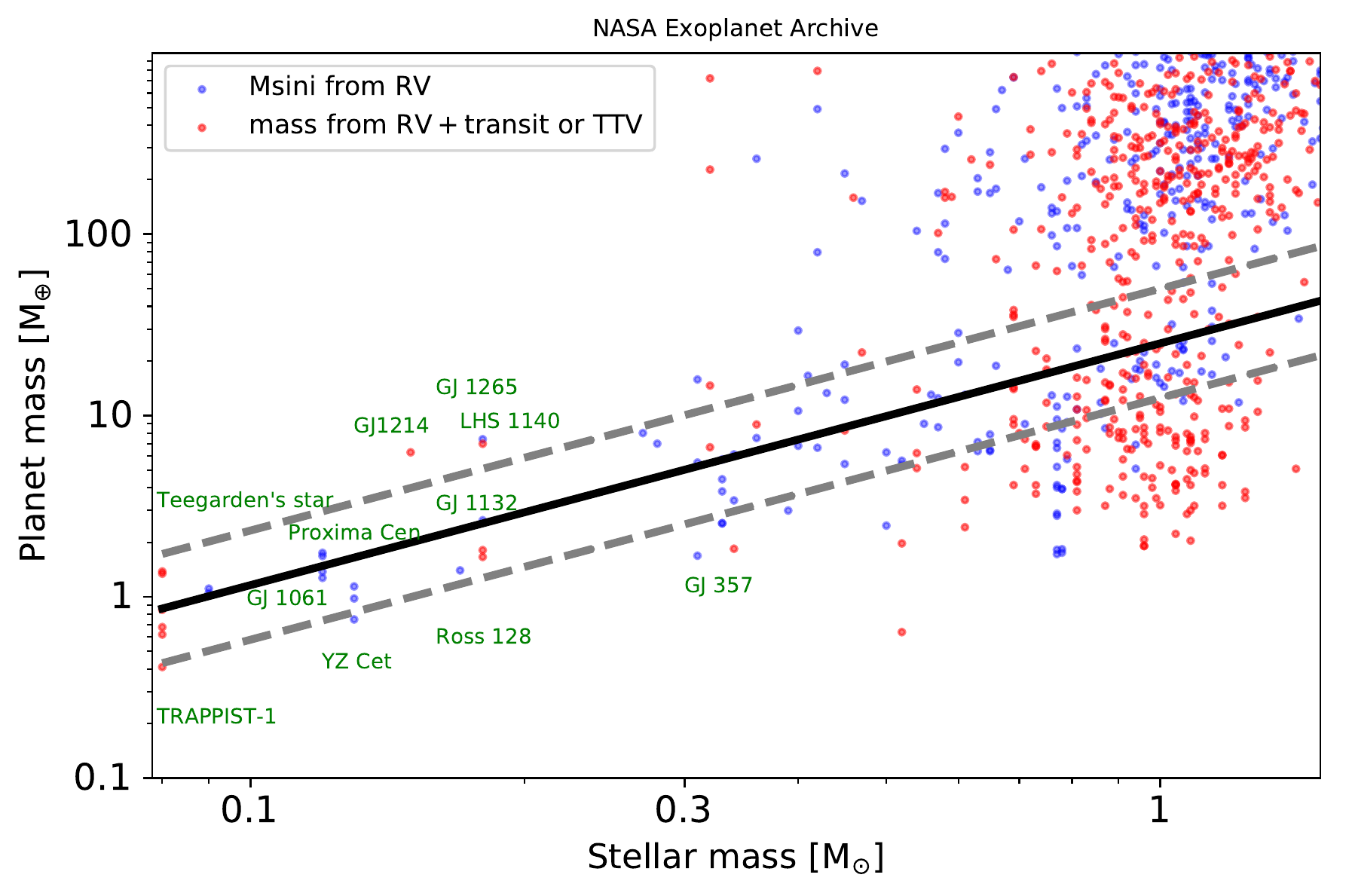}
       \caption{  
       Plot of the observed planet mass vs the stellar mass. The blue dots are the planets only detected by radial velocity surveys with a low mass limit,  and the red dots are the planets with true  masses  either  from a combined radial velocity and transit surveys, or from transit timing variation measurements.   The black line is adopted from \eq{M_iso_vis}, indicating the pebble isolation mass (see details in \se{obs}), and the grey lines show a factor of two variation. The masses of planets seem to correlate with the masses of their central stars.    The names of typically planetary systems around very low-mass stars are labeled in green,  including $\rm TRAPPIST$-$1$, \citep{Gillon2016}, $\rm Proxima \ Centauri$ \citep{Anglada-Escude2016}, $\rm Ross\ 128$ \citep{Bonfils2018}, $\rm LHS\ 1140$ \citep{Dittmann2017}, $\rm YZ \ Cet$ \citep{Astudillo-Defru2017}, $\rm Teegarden's  \ star$ \citep{Zechmeister2019}, $\rm GJ \ 1214$  \citep{Charbonneau2009}, $\rm  GJ \ 1265$ \citep{Luque2018}, $\rm GJ \ 1132$ \citep{Berta-Thompson2015}, $\rm GJ \ 1061$ \citep{Dreizler2019} and $\rm GJ \ 357$ \citep{Luque2019}. 
    }
\label{fig:massobs}
\end{figure} 

In order to get some clues, as a first attempt, here we plot the masses of the observed exoplanets as a function of their stellar masses in \fg{massobs}. The data are adopted from NASA exoplanet archive (https://exoplanetarchive.ipac.caltech.edu/) in August, 2019. Only the planets with known masses are selected and we further separate these planets into two samples.  The masses of the planets measured only by radial velocity (RV)  are shown in blue, while those measured by a combined RV and transit, or by transit timing variation (TTV) are shown in  red.  A low mass limit is inferred for the former sample while the true mass is able to obtain for the latter one.  The black line represents a linear mass correlation between the planets and their stellar hosts and the grey dashed lines mark a factor of two variation (see physical motivation and discussion in \se{obs}). We can see that the observed small planets around low-mass stars ($M_{\star} \lesssim 0.3 \Ms$) actually follow this linear trend well. Clearly, there is no  massive planets detected around very-low mass stars.  It is worth pointing out that  although the detection probability varies with different surveys,  for the same planet, both RV and transit signals are generally easier to detect when it orbits around GK stars than less massive M dwarfs. The detection limit is  towards to smaller planets around more massive stars. Therefore, the observed linear mass trend between the planets and stars is not due to the observational bias.

The key is to understand which physical process cause the planets to end up with a certain mass and how it varies with the masses of stellar hosts. 
  The population synthesis approach is an ideal tool to explore this issue.  By construction, population synthesis interprets  the complex physical processes into simplified recipes. Each recipe can be obtained from more sophisticated simulations that focus on individual processes. In this way, a deterministic planet formation model can be built.  The input parameters are connected to the planetary birth conditions, such as the disk gas accretion rate, the dust-to-gas ratio and the disk lifetime.  A large number of synthetic planets can be generated by randomlizing their initial conditions over appropriate distribution functions. The influence of each physical processes on the final planet populations can be further analysed and statistically compared with the observed exoplanets (see \citealt{Benz2014} for a review). 
 
  \cite{Ida2004a} first used the population synthesis approach to study the planet formation around solar-mass stars, and later \cite{Ida2005} extended their work to stars of different masses. They found that the growth of Jupiter-like planets are suppressed around M dwarf stars and the formation rate of gas giants  increases with their stellar masses. This prediction was later confirmed by RV results \citep{Johnson2007,Johnson2010}.
  Based on \cite{Mordasini2009}'s population synthesis model,  \cite{Alibert2011} stated that the forming planets are crucially related with their initial disk mass and gas disk lifetime, which vary with the stellar type.   \cite{Coleman2016} pointed out that  the size of the accreted planetesimals also influences the final planetary masses.  

All the mentioned studies are nevertheless based on the planetesimal accretion scenario \citep{Safronov1972,Wetherill1989,Ida1993,Kokubo1998}. The planetesimals are kilometer or larger in size, and their growth is dominated by the gravitational force. The aerodynamic drag force from the disk gas is negligible during their mutual interactions and encounters. On the other hand, \cite{Ormel2010} and \cite{Lambrechts2012} proposed that smaller particles affected by gas drag are much efficiently accreted by the planet when they drift and cross the orbit of the planet.  The accretion criterion is that particles should be affected by gas drag on a similar timescale as their gravitational scattering by the planet - such particles have a characteristic size of millimieter-centimeter. This accretion process is also termed as pebble accretion (see \citealt{Johansen2017,Ormel2017} for reviews).   \cite{Lambrechts2014b} found that pebble accretion terminates when the planet reaches the pebble isolation mass. This happens when the gravity of the planet is strong enough to open a gap and reverse the local disk pressure gradient. Such a process stops the inward drifting pebbles, and therefore the following mass growth by pebble accretion. The pebble isolation mass for planets around solar-mass stars is roughly $10$$-$$20 \Me$ \citep{Lambrechts2014b,Bitsch2018,Ataiee2018}, and it linearly scales with the masses of their host stars.  

 \cite{Bitsch2015} considered the planetary core growth by pebble accretion and studied the formation of planets at different disk locations and evolutional stages. \cite{Johansen2018} investigated under which disk conditions  the planet growth can overcome the disk migration. 
\cite{Lambrechts2019} suggested that whether a system ends up with low-mass, terrestrial-planets or super-Earth planets could  be determined by the disk pebble mass flux (reservoir of pebbles in the disk). All these studies nevertheless focus on the solar-mass stars. 
 
\cite{Ormel2017}  proposed a scenario for the formation of TRAPPIST-$1$ and other compact systems around very low-mass M-dwarf stars, where  embryos are only originated at the water ice line. They suggested  that all seven planets of roughly earth-mass around TRAPPIST-$1$ star  could be an indication of planet masses set by the pebble isolation mass.  The followed-up numerical simulations by \cite{Schoonenberg2019} found that these forming planets would contain $\equiv$$10\%$ water in mass. 
  This fraction is different from the scenarios that assume the initial distribution of embryos is more radially spread  (\eg , see \cite{Ogihara2009} for low-mass stars and  \cite{Izidoro2019} for solar-mass stars).  There is lack of such pebble accretion driven models that study how planets form around stars of different masses and at different disk locations.

In this paper, we will construct a pebble-driven planet population synthesis model and focus on the planet formation around stars of various masses, ranging from very low-mass M dwarfs ($0.08 \Ms$) to the Sun-like stars ($1 \Ms$). The fundamental question we aim to address in this work is how and why the planet masses correlate with the masses of their stellar hosts. Equivalently, we seek for the key planet formation process that determines the characteristic mass of the planets.

This paper is organized as follows. In \se{method} we describe the base of our model. In \se{result} we present the illustrated simulations, and discuss the influence of key parameters on planet growth and migration.  We focus  on the final masses and water contents of the resultant planets as functions of their birth locations and time in \se{map}. The synthetic planet population generated by a Monte Carlo approach are compared with the observed exoplanets in \se{obs}.  \Se{conclusion} is the discussion and conclusion.

\section{Method}
\label{sec:method}

In this section we describe the adopted disk model in \se{disk},  planet mass growth by pebble and gas accretion in \se{growth}, and planet migration in \se{migration}.

\subsection{Disk model}
\label{sec:disk}

We use a standard $\alpha$ viscosity prescription  \citep{Shakura1973}  for the disk gas angular momentum transfer, $\nu = \alpha_{\rm \nu} c_{\rm s} H_{\rm g}$, where $\nu$ is the viscosity, the viscous coefficient $\alpha_{\rm \nu}$ represents  the efficiency of global gas angular momentum transfer, $H_{\rm g }$ is the scale height of the gas disk, $c_{\rm s}= H_{\rm g} \Omega_{\rm K}$ is the gas sound speed and  $\Omega_{\rm K}$ is the Keplerian angular frequency.  For a steady state  viscous accretion disk, the mass accretion rate through the inner disk regions is related to the viscosity and the gas surface density $\Sigma_{\rm g}$ through 
\begin{equation}
\dot M_{\rm g} =  3 \pi \nu \Sigma_{\rm g}.
\label{eq:steady}
\end{equation}
 This indicates that  the viscosity can be quantified by  the gas disk accretion rate $ \dot M_{\rm g}$ and the gas surface density $\Sigma_{\rm g}$. 
The measured  $\dot M_{\rm g}$ and inferred $\Sigma_{\rm g}$ from disk observations are consistent with   $\alpha_{\rm \nu}$ in the broad range of $10^{-2}-10^{-3}$  \citep{Hartmann1998,Andrews2009}. This value  generally agrees with the turbulence driven by the magnetorotational instability  (MRI)  in ideal magetohydynamical (MHD)  simulations \citep{Balbus1998}. In this study we assume a classical viscous accretion driven disk model,  which does not account for MHD disk winds and other non-ideal MHD effects \citep{Bai2013,Gressel2015}. Here $\alpha_{\rm \nu} =10^{-2}$ is adopted as our fiducial value.

   \subsubsection{$\dot M_{\rm g}$ evolution}
 The  disk angular momentum transfer (decline of $ \dot M_{\rm g}$) follows two processes: internal viscous accretion onto the central star and evaporation away to the interstellar medium by high energetic stellar UV and X-ray photons.  
   When the viscous accretion dominates the gas removal, we adopt the self-similar solution from \cite{Lynden-Bell1974,Hartmann1998},
   \begin{equation}
    \dot M_{\rm g} = \dot M_{\rm g0}
    \left[  1 + \frac{t}{t_{\rm vis}} \right]^{-\gamma}.
      \label{eq:mdot_acc}
   \end{equation}
where $\dot M_{\rm g0}$ is the initial disk accretion rate,  $\gamma= (5/2 + s) / (2+s) $  and  $s $ is the gas surface density gradient. 
 
The characteristic viscous accretion timescale is given by  
     \begin{equation}
    t_{\rm vis} = \frac{1}{3(2+s)^2} 
    \frac{r_{\rm d0}^2}{\nu_{0}},
    \label{eq:tvis}
   \end{equation}
   where $r_{\rm d0}$ and  $\nu_{0}$ are the initial characteristic size of the gas disk and the viscosity at that radius.  The gas flux changes sign at a radius  
       \begin{equation}
    r_{\rm c} = r_{\rm d0} \left[ \frac{1 + t/t_{\rm vis} }{ 2(2 +s)}    \right] ^{1/(2 + s)}.
   \end{equation}
Note that to conserve the disk angular momentum,  gas moves inward (accretion) when  $r<r_{\rm c}$ and moves outward when $r>r_{\rm c}$ (spreading).
  The outer edge of the disk expands at the same rate as  $r_{\rm c}$, and we assume the characteristic size $r_{\rm d} \simeq 2 r_{\rm c}$ for the adopted disk profile.  Observationally, \cite{Tazzari2017} found that Lupus disks are larger and less massive than younger Taurus-Auriga and Ophiuchus disks, which might support the assumption of viscous expanding disks.

  We also take into account the effect of angular momentum transportation by stellar photoevaporation. In this mechanism, the surface layer of disk is heated up to $10^{3} $ K by high-energy photons from the central star. The gas becomes unbound and flows away from the disk as a wind (see \cite{Alexander2014} for a review). When the photoevaporation rate is higher than the viscously driven accretion rate, photoevaporation dominates the disk dispersal. In this regime the accretion rate is modified as  
    \begin{equation}
    \dot M_{\rm g} = \dot M_{\rm g0}
    \left[  1 + \frac{t}{t_{\rm vis}} \right]^{-\gamma} \exp \left[ -\frac{t-t_{\rm pho}}{\tau_{\rm pho} }\right],
    \label{eq:mdot_pho}
   \end{equation}
   where the exponential decay term corresponds to the mass lost from the photoevaporation, characterized with a dispersal timescale $\tau_{\rm pho}$.
   Here we only focus on the X-ray driven photoevaporation. The critical accretion rate is given by \cite{Owen2012},    
    \begin{equation}
    \begin{split}
    \dot M_{\rm pho} & =  6 \times 10^{-9}
    \left(\frac{M_{\star}}{1 \ M_{\odot}} \right)^{-0.07} 
    \left(\frac{L_{\rm X}}{30 \ \rm erg \ s^{-1}} \right)^{1.14} 
    \rm \ M_{\odot}\, yr^{-1}\\
    & = 6 \times 10^{-9}  
    \left(\frac{M_{\star}}{1 \ M_{\odot}} \right)^{1.6}
    \rm \ M_{\odot} \, yr^{-1}, 
     \label{eq:pho}
   \end{split}
\end{equation}
 The latter equality in \eq{pho} follows the stellar X-ray luminosity and mass correlation, $L_{\rm  X} \propto M_{\star}^{1.5}$ \citep{Preibisch2005a,Gudel2007}. The  young stars are magnetic active and maintain their energetic X-ray radiations during the evolution of the pre-main sequences  \citep{Preibisch2005b}.
   The onset of the photoevaporation  $t_{\rm pho}$ is obtained  by replacing \eq{pho} in the left-hand side of  \eq{mdot_acc}. 
   The gas removal timescale by the stellar photoevaporation is calculated as    
   \begin{equation}
    \tau_{\rm pho}= M_{\rm d}(t= t_{\rm pho}) / \dot M_{\rm pho},
    \end{equation}
 where the gas disk mass $  M_{\rm d}(t)= \int_0^{r_{\rm d}} {2\pi r \Sigma_{\rm g}}(r,t) {\rm d}r$.
    
    Disk observations indicate that  the typical disk lifetime lasts for a few Myr \citep{Haisch2001,Mamajek2009}, 
    while final depletion of disk gas is much faster, inferred to be shorter than $1$  Myr \citep{Luhman2010,Williams2011}. 
As we will show in \se{result},  these features are consistent with our adopted viscous accretion + stellar evaporation disk model.

  \subsubsection{Two-component disk structure}

Based on the quasi-steady state assumption in \eq{steady}, we can calculate gas surface density and disk temperature at each $\dot M_{\rm g}$.  A two-component disk structure is constructed based on different heating mechanisms \citep{Garaud2007}.  In the inner disk the dominated heating energy is generated by the internal viscous dissipation \citep{Ruden1986}, whereas in the outer disk it is govern by the irradiation from the central star onto the surface layer of the disk \citep{Chiang1997}. 
We assume that the inner viscously heated disk is optically thick, where the disk opacity $ \kappa = \kappa_0 \left( T_{\rm g}/1\K \right) \rm \ g\ cm^{-2} $ with the coefficient $\kappa_0=0.01$  for the fiducial value. The above prescription assumes small grains dominate the dust opacity and neglects the more complicated temperature dependence due to sublimation of different species \citep{Bell1994}. Noteworthy, $ \kappa$ is expected to be correlated with the disk metallicity, which will be further discussed in \se{obs}.  
 The outer disk region heated by  stellar irradiation is assumed to be vertically optically thin for simplicity.

For the inner viscously heated disk, the gas surface density, disk temperature and gas disk aspect ratio  ($h_{\rm g} =H_{\rm g }/r$) are given by  
\begin{equation}
\begin{split}
    \Sigma_{\rm g,vis} = &  132
    \left( \frac{\dot M_{\rm g}}{10^{-8} \Msyr} \right)^{1/2}  \left(\frac{M_{\star}}{1 \ M_{\odot}} \right)^{1/8}   
    \left(\frac{\alphag} {10^{-2}} \right)^{-3/4}\\
    & \left(\frac{\kappa_0}{10^{-2}} \right)^{-1/4} 
    \left(\frac{r}{1 \AU} \right)^{-3/8}
    \ \rm g \ cm^{-2},
    \label{eq:sig_vis}
    \end{split}
    \end{equation}
\begin{equation}
\begin{split}
T_{\rm g,vis} = & 280
\left( \frac{\dot M_{\rm g}}{10^{-8} \Msyr} \right)^{1/2}
\left(\frac{M_{\star}}{1 \ M_{\odot}} \right)^{3/8} 
\left(\frac{\alphag} {10^{-2}} \right)^{-1/4}\\
& \left(\frac{\kappa_0}{10^{-2}} \right)^{1/4} 
\left(\frac{r}{1 \AU} \right)^{-9/8}
\ \rm K,
  \end{split}
\end{equation}
\begin{equation}
\begin{split}
h_{\rm g,vis} = & 0.034
\left( \frac{\dot M_{\rm g}}{10^{-8} \Msyr} \right)^{1/4}
\left(\frac{M_{\star}}{1 \ M_{\odot}} \right)^{-5/16} \\
&  \left(\frac{\alphag} {10^{-2}} \right)^{-1/8}
\left(\frac{\kappa_0}{10^{-2}} \right)^{1/8}
\left(\frac{r}{1 \AU} \right)^{-1/16},
\label{eq:h_vis}
  \end{split}
\end{equation}
where $M_{\star}$ is the mass of the central star and $r$ is the disk radial distance. 
The detailed derivation for the above expressions are given in \ap{derivation}. 
For the stellar irradiated disk, the surface density, temperature and gas aspect ratio are given by \citep{Ida2016}    
\begin{equation}
  \begin{split}
\Sigma_{\rm g,irr} = & 250
\left( \frac{\dot M_{\rm g}}{10^{-8} \rm \Msyr} \right)
\left(\frac{M_{\star}}{1 \ M_{\odot}} \right)^{9/14} 
\left(\frac{L_{\star}}{1 \ L_{\odot}} \right)^{-2/7} \\
& \left(\frac{\alphag} {10^{-2}} \right)^{-1}
\left(\frac{r}{1 \AU} \right)^{-15/14}
\ \rm g \ cm^{-2},
\label{eq:sig_irr}
  \end{split}
\end{equation}
\begin{equation}
  \begin{split}
    T_{\rm g,irr} = & 150
    \left(\frac{M_{\star}}{1 \ M_{\odot}} \right)^{-1/7} 
    \left(\frac{L_{\star}}{1 \ L_{\odot}} \right)^{2/7}  
     \left(\frac{r}{1 \AU} \right)^{-3/7}
    \ \rm K,
    \label{eq:T_irr}
      \end{split}
\end{equation}
\begin{equation}
    h_{\rm g,irr} = 0.0245
    \left(\frac{M_{\star}}{1 \ M_{\odot}} \right)^{-4/7} 
    \left(\frac{L_{\star}}{1 \ L_{\odot}} \right)^{1/7} 
    \left(\frac{r}{1 \AU} \right)^{2/7}.
    \label{eq:h_irr}
\end{equation}
To sumarrize, the above disk profile can be  expressed as   $\Sigma_{\rm g} = {\Sigma_{\rm g0} (r/\AU)}^{s}$, $T_{\rm g} = {T_{\rm g0} (r/\AU)}^{p}$ and $h_{\rm g} = {h_{\rm g0} (r/\AU)}^{q}$ where the subscript $0$ means the value at $1$ AU. In the viscously heated region,  $s=-3/8$, $p=-9/8 $ and $q = -1/16$, whereas in the stellar irradiated region $s=-15/14$, $p=-3/7 $ and $q = 2/7$. 
Given by the expression of $\dot M_{\rm g}(t)$ in \eqs{mdot_acc}{mdot_pho}, we can calculate $\Sigma_{\rm g}(t)$, $T_{\rm g}(t)$ and $h_{\rm g}(t)$  accordingly. 

The transition radius in the disk that separates the viscously heated region from the stellar heated region  is expressed as  
\begin{equation}
      \begin{split}
    r_{\rm tran} = & 2.5
    \left( \frac{\dot M_{\rm g}}{10^{-8} \rm \Msyr} \right)^{28/39}
    \left(\frac{M_{\star}}{1 \ M_{\odot}} \right)^{29/39} \\
  &  \left(\frac{L_{\star}}{1 \ L_{\odot}} \right)^{-16/39} 
     \left(\frac{\alphag} {10^{-2}} \right)^{-14/39}
    \left(\frac {\kappa_0}{10^{-2}} \right)^{14/39}
   \AU.
     \end{split}
     \label{eq:rtrans}
\end{equation}
We note that the corresponding length scale for viscous disk evolution is $R_{\rm d0}$.  Since the characteristic disk size is generally much larger than the transition radius ($R_{\rm d0} \gg r_{\rm tran}$), we simply adopt $s =-15/14$ from the outer stellar irradiation region for the evolution of $\dot M_{\rm g}$ in \eqs{mdot_acc}{mdot_pho}.   

\subsubsection{The water ice line}
In protoplanertary disks $\rm H_2O$ vapor condenses into ice (ice sublimates into vapor) when the water vapor pressure is higher (lower) than its saturated vapor pressure. The water ice line corresponds to the equilibrium point where the sublimation rate of  $\rm H_2 O$ ice equals the condensation  rate of the vapor. Interior to the ice line location $r_{\rm ice}$, $\rm H_2 O$ is in the form of vapor,  while exterior to $r_{\rm ice}$ it exists as the ice.    
The $\rm H_2 O$ saturated pressure determines the maximum amount of allowed water in form of vapor. Laboratory experiments fit it as an exponential decay with the disk temperature, which is given by \cite{Haynes1992}, 
 \begin{equation}
 P_{\rm sat} = P_0 \rm \exp(-T_0/T_{\rm g})
  \label{eq:Psat}
  \end{equation}
  where $P_0= 6.034 \times 10^{12} \rm \  g \ cm^{-1} \ s^{-2}$, $T_0=5938 \ \rm K$. 
  The corresponding water vapor pressure is  
   \begin{equation}
 P_{\rm H_2 O} = \frac{\rho_{\rm H_2 O}}{m_{\rm H_2O}} k_{\rm B} T_{\rm g} = \frac{Z_{\rm H_2O } \Sigma_{\rm g}}{\sqrt{2 \pi} H_{\rm g} m_{\rm H_2 O}} k_{\rm B} T_{\rm g} ,
  \label{eq:P_h2o}
  \end{equation}
 where $m_{\rm H_2O}$ is the molecule weight of $\rm H_2O$, $k_{\rm B} = 1.38 \times 10^{-16} \rm\ erg \ K^{-1} $ is the Boltzmann constant and the water metallicity approximates to the disk metallicity, $Z_{\rm H_2 O} \approx Z_{\rm d}$.  
 The location of the water ice line is calculated when these two are equal,  
  \begin{equation}
   P_{\rm sat}(r_{\rm ice})  = P_{\rm H_2 O} (r_{\rm ice}).
     \label{eq:P_eq}
   \end{equation}
   It is worth pointing out that the water ice line is not always at one fixed disk temperature since it is also related with the disk density (\eq{P_h2o}).  However, the saturated pressure decreases rapidly with temperature. The disk temperature at the ice line therefore only modestly varies from  $150$ to $190 \K$ for the relevant protoplanetary disk conditions. The corresponding ice line location $r_{\rm ice}$ is in the viscously heated region in the early phase when $\dot M_{\rm g}$ is high, and moves into the stellar irradiated region in the late phase when $\dot M_{\rm g}$ becomces low. For a simple analytical purpose, we assume $T_{\rm ice} = 170 \rm \ K$, and the ice line can be derived separately as  
     \begin{equation}
     \begin{split}
    r_{\rm ice,vis} = & 1.56
    \left( \frac{\dot M_{\rm g}}{10^{-8} \rm \  M_{\odot}yr^{-1}} \right)^{4/9}
    \left(\frac{M_{\star}}{1 \ M_{\odot}} \right)^{1/3} \\
 &   \left(\frac{\alphag} {10^{-2}} \right)^{-2/9}
    \left(\frac {\kappa_0}{10^{-2}} \right)^{2/9}
   \AU,
   \label{eq:rice_vis}
        \end{split}
\end{equation}
and 
\begin{equation}
\begin{split}
    r_{\rm ice,irr} = & 0.75 \left(\frac{M_{\star}}{1 \ M_{\odot}} \right)^{-1/3} 
    \left(\frac{L_{\star}}{1 \ L_{\odot}} \right)^{2/3} 
   \AU.
   \label{eq:rice_irr}
   \end{split}
   \end{equation}
   Our \eq{rice_vis}, the ice line location in the viscously heated disk, is in agreement with Eq.(2) of \cite{Mulders2015b}.
   The ice line for this two-component disk is therefore given by $r_{\rm ice} = \max{(r_{\rm ice,vis}, r_{\rm ice,irr} )}$.

\cite{Lodders2003} obtained an equal silicate-to-water mass ratio for the protosolar element abundances under the chemical equilibrium assumption that all the carbon atoms form methane ($\rm CH_{\rm 4}$). This means, besides forming silicate minerals (\eg , olivine and pyroxene), all the residual oxygen atoms react to form $\rm H_{2}O$.  However,  for the typical pressure and temperature condition in protoplanetary disks, carbon instead resides in the molecules  $\rm CO$ and $\rm CO_{2}$ due to the non-equilibrium chemistry \citep{Woitke2009,Henning2013}. Thus the amount of oxygen that  forms $\rm H_2O$ may be significantly less compared to \cite {Lodders2003}'s calculation. The actual  silicate-to-water mass ratio depends on abundance of carbon-bearing species (\eg , refractory carbon, $\rm CO$ and $\rm CO_{2}$) in protoplanetary disks, relying on the complex disk chemistry.

From the observational perspective, \cite{Gail2017} summarized from the  literature on comet Halley  that the most of solid carbon in the comet forming regions resides in the so-called refractory carbonaceous materials (hydrocarbons and complex organic matter) whose the condensation temperature is around $ 400 \K$.  
 \cite{Fulle2017} reported a high dust-to-ice ratio in comet $\rm 67P$ and further deduced that  $50\%$ of its volume is dominated by such carbonaceous materials while the silicate and the ice contribute equally to the rest of the volume. However, the meteorites whose parent bodies originated in the terrestrial planets region and the asteroid belt are found to be depleted in such carbonaceous materials \citep{Wasson1988,Bergin2015}. The Earth is extremely  carbon-deficient compared to the solar abundance, even a large fraction of carbon may be hidden in the Earth's core \citep{Allegre2001,Marty2012}.  How the depletion of the refractory carbon occurred in the inner region of the solar system is still an unsolved issue.   

In order to account for the composition of the forming planets,
\cite{Bitsch2019} derived a water mass fraction of $\simeq 35\%$ in the region between the $\rm H_2O$ and $\rm CO_2$ condensation lines ($70 \K \lesssim  T \lesssim 170 \K$). They adopted the chemical species model from \cite{Madhusudhan2014}, based on the detailed radiation and thermo-chemical disk modelling \citep{Woitke2009} and the solar chemical composition \citep{Asplund2009}. The refractory carbonaceous materials are not taken into account in their chemical model.
 This $35\%$ of  the water content is more close to the values measured  from carbonaceous chondrites \citep{Garenne2014,Braukmuller2018}.  In our model the planet growth and migration mainly occur inside of the $\rm CO_{2}$ condensation line.  We therefore assume that the pebbles exterior to the water ice line contain $35\%$ water ice and $65\%$ silicate.  When pebbles drift interior to the water ice line,  the water ice sublimates and only the silicate component remains as solids.

\subsection{Planet growth}
\label{sec:growth}
Here we describe the planetary core growth by pebble accretion in \se{pebble}  and envelope growth by gas accretion in \se{gas}, respectively. 

\subsubsection{Pebble accretion}
\label{sec:pebble}

We start with the protoplanetary embryo of $10^{-2} \Me$ and allow them to grow core masses by  pebble accretion. 
The pebble accretion efficiency refers to the fraction of pebbles accreted by the planet when they drift across the planet's orbit. In the $2$D regime it reads \citep{Liu2018}
 \begin{equation}
\begin{split}
\varepsilon_\mathrm{PA,2D} & =  \frac{0.32}{\eta } \sqrt{ \frac{M_{\rm p} }{M_\star}  \frac{\Delta v}{ v_{\rm K} } \frac{1}{\taus}}  = 3\times 10^{-3}  \left( \frac{ M_{\rm p}}{0.01 \Me} \right)^{2/3} \\
&  \left( \frac{ \taus}{10^{-2}} \right)^{-1/3}   \left( \frac{\eta}{3 \times 10^{-3}}  \right)^{-1}  
\label{eq:eps-2D}
\end{split}
\end{equation}
where $M_{\rm p}$ and $M_{\star}$ are the masses of the  planet and the star, $\Delta v$ and $V_{\rm K}$ are the relative velocity between the planet and the pebble, and the Keplerian velocity at the planet location, $\taus $ is the Stokes number of the pebble measuring its aerodynamical size, $\eta = -h_{\rm g}^2 ({\partial \rm ln P/  \partial ln r }) /2=  (2-s-q) h_{\rm g}^2/2$,  and $P$ is the gas pressure in the disk.  The latter expression in \eq{eps-2D} assumes that the planet is in the shear regime where the relative velocity is equal to the shear velocity between the planet and the pebble.  

The pebble accretion efficiency in the $3$D regime reads \citep{Ormel2018}
\begin{equation}
\label{eq:eps-3D}
\begin{split}
\varepsilon_\mathrm{PA, 3D} & =  \frac{0.39 }{\eta h_\mathrm{peb}} \frac{M_{\rm p}}{ M_\star}  =  4 \times 10^{-3}      \left( \frac{ M_{\rm p}}{0.01 \Me} \right)    \left( \frac{ M_{\star}}{1 \ M_{\odot}} \right)^{-1}  \\
&  \left( \frac{ h_{\rm peb}}{ 3 \times 10^{-3}} \right)^{-1} 
   \left( \frac{\eta}{3 \times 10^{-3}}  \right)^{-1}.
 \end{split}
\end{equation}
The total pebble accretion efficiency is given by $\varepsilon_\mathrm{PA} = \sqrt{\varepsilon_\mathrm{PA, 3D}^{-2} + \varepsilon_\mathrm{PA, 2D}^{-2}}$. The above efficiency formulas are valid for particles whose Stokes numbers are  comparable or lower than unity. Larger Stokes number particles are less gas-aided and their aerodynamical behavior is more like planetesimals. We neglect the accretion in that regime and set  $\varepsilon_\mathrm{PA} =0$ when $\taus>10$.

Whether pebble accretion is in $2$D or $3$D is determined by the ratio between the pebble accretion radius and the vertical layer of pebbles \citep{Morbidelli2015}. 
  The pebble scale height is given by \citep{Youdin2007}, 
 \begin{equation}
 H_{\rm peb} = \sqrt{ \frac{\alphat }{\alphat + \taus} } H_{\rm g}.
 \label{eq:hpeb}
\end{equation}

Here $\alphat$ is the dimensionless parameter that measures the turbulent diffusivity. The turbulent diffusion coefficient approximates the same as the  local turbulent viscosity when the disk turbulence is driven by magnetorotational instability \citep{Johansen2005,Zhu2015,Yang2017}. Hereafter we do not distinguish these two and also refer $\alphat$ as the local turbulent viscosity strength. In reality $\alphat$ can differ from the averaged, global disk angular momentum transport efficiency $\alphag$ due to layered accretion \citep{Fleming2003,Turner2008}.  
 This adopted turbulent $\alphat$ is more relevant to the local planet formation scale. For instance, it determines the processes such as gap-opening in the vicinity of the planet, the diffusion of gas and dust particles across the gap.

The disk turbulent strength can be inferred from the comparison between modelling the dust settling and  the observed disk morphology. \cite{Pinte2016} suggested $\alphat \simeq10^{-3}$ to $ 10^{-4}$ for the HL tau disk by Atacama Large Millimeter Array (ALMA) observation. \cite{Ovelar2016} obtained $\alphat \simeq 10^{-3}$ from a combined ALMA and Spectro-Polarimetric High-contrast Exoplanet Research (SPHERE) data for the transitional disks. Measuring  the non-thermal motion of the disk gas from $\rm CO$ line emission provides an upper limit of $\alphat$. \cite{Flaherty2015,Flaherty2017} found  that the turbulent velocity should be smaller than a few percent of the sound speed in $\rm HD  \ 163296$ disk, indicating $\alphat \lesssim 10^{-3}$. Since the disk turbulent level is not well constrained, we adopt both $\alphat = 10^{-3}$ and $10^{-4}$ for our analysis.  The influence  of varying this parameter on planet formation will be further demonstrated and discussed in \ses{map}{obs}.

The gas and the pebble mass fluxes are  given by $\dot M_{\rm g} = 2\pi r \Sigma_{\rm g} v_{\rm g} $ and $\dot M_{\rm peb} = 2\pi r  \Sigma_{\rm peb} v_{\rm peb}$ where $v_{\rm g} $ and  $v_{\rm peb}$ are the radial velocities of the gas and pebbles, respectively. The local metallicity, \ie , the surface density ratio between pebbles and the gas is given by  
\begin{equation}
\begin{split}
Z_{\rm p} = & \frac{\Sigma_{\rm peb}}{\Sigma_{\rm g}}  = \frac{ \dot M_{\rm peb}  }{\dot M_{\rm g} } \frac{1}{1 + v_{\rm peb}/v_{\rm g} }
=  \xi / \left[ \frac{2(2-s-q) }{3} \frac{\taus}{\alphag }+1 \right],
\label{eq:sig_ratio}
\end{split}
\end{equation}
where $\xi \equiv \dot M_{\rm peb} /\dot M_{\rm g}$ \citep{Ida2016} represents the mass flux ratio between the gas and pebbles.
The first term in the brackets of \eq{sig_ratio} represents the velocity ratio between the pebble and gas, which scales with $\taus/\alphag$.
In the limit of small pebbles (\eg , $\lesssim$ mm-sized particles) that are well-coupled to the disk gas ($\taus \lesssim \alphag$),   $Z_{\rm p}$  remains the same as the initial disk metallicity $Z_{\rm d}$.  In this case gas and pebbles drift inward at the same radial speed, and $\xi$ is  a good proxy of the disk metallicity.

The mass accretion rate onto the planet is given by
\begin{equation}
\dot M_{\rm PA} = \varepsilon_\mathrm{PA} \dot M_{\rm peb} =  \varepsilon_\mathrm{PA} \xi \dot M_{\rm g}.
\label{eq:m_PA}
\end{equation}
From \eqss{eps-2D}{eps-3D}{sig_ratio} the growth by pebble accretion increases with the planet mass, the local metallicity and the disk accretion rate \footnote{  Note that we neglect the midplane gas radial velocity for calculating $\varepsilon_\mathrm{PA}$ based on \eqs{eps-2D}{eps-3D} due to the layered accretion disk assumption.  However, for completeness, we also conduct simulations of $\varepsilon_\mathrm{PA}$  with the gas radial velocity term included,  which has little effect on our findings (see \ap{lowalpha}).
}.

In this paper, we neglect the pebble growth and treat the pebbles to be all one-millimeter in size. This characteristic size is  motivated from the measured spectral index from the disk observations at millimeter wavelengths \citep{Draine2006,Perez2015}. In addition, based on the laboratory experiments on collisions between silicate dust aggregates, \cite{Zsom2010} found that their growth is limited to millimeter sizes due to bouncing. \cite{Musiolik2019}  suggested that ice aggregations have a similar growth pattern as silicates at low disk temperature ($T_{\rm g}\lesssim 180\K$). Note that in this circumstance, pebbles stall at the sizes before they reach the fragmentation regime \citep{Guttler2010}.  Their masses only reduce by $35\%$ when they across the water ice line. The change in size is even minor ($\lesssim 10\%$). We neglect this pebble size change cross the ice line. 

 Since the planets would migrate towards and accrete most of their masses around the transition radius (see \se{type_I}), the particle size at $r_{\rm tran}$ is an important quantity, which can be derived from \eqs{sig_vis}{rtrans} as,  
    \begin{equation}
      \begin{split}
  R_{\rm peb}   & = \frac{ \taus \Sigma_{\rm g} }{ \sqrt{2\pi} \rho_{\bullet} } \\
  &= 2.5  \rm \  mm \left( \frac{\taus}{0.01} \right) \left(\frac{M_{\star}}{1 \ M_{\odot}} \right)^{-0.2}  \left( \frac{\dot M_{\rm g}}{10^{-8} \Msyr} \right)^{0.2}   \\
    & \left(\frac{L_{\star}}{1 \ L_{\odot}} \right)^{0.2} \left( \frac{\rho_{\bullet}}{1.5 \rm g\,cm^{-1}} \right)^{-1}  
    \left(\frac{\alphag} {10^{-2}} \right)^{-0.6}
     \left(\frac{\kappa_0}{10^{-2}} \right)^{-0.4},
    \end{split}
    \label{eq:R_peb}
    \end{equation}
    where $ \rho_{\bullet}$ is the internal density of the planet.  For the typically adopted disk parameters in \eq{R_peb} and its weak dependence on $\dot M_{\rm g}$, $M_{\star}$ and $L_{\star}$, these mm-sized pebbles correspond to Stokes numbers $\taus$ of $\approx$$0.01$, both for disks around solar-mass stars as well as around $0.1\Ms$ stars.

 However, $\dot M_{\rm g}$ drops significantly during the late phase of the disk evolution when the stellar photoevaporation commences. At that time even mm-sized particles have larger Stokes numbers, and hence decouple from the motion of the gas. Therefore, pebbles drift inwardly faster than the disk gas.  On the other hand, the disk becomes more dusty because of the gas removal.  In the early  phase $\xi= \xi_0$ and in the late stellar photoevaporation phase $\xi= \xi_0 \dot M_{\rm pho}/\dot M_{\rm g}$, where $\xi_0$ is the initial pebble to gas flux ratio and $\dot M_{\rm pho}$ is the onset accretion rate when photoevaporation dominates.  Pebble accretion benefits from this enhanced metallicity.  Nonetheless, stellar photoevaporation  occurs late and dominates for a relatively short timespan. It is still a justified approximation that the flux ratio $\xi$ is a good indicator of the overall metallicity $Z_{\rm d}$ in protoplanetary disks.

When the embedded planet grows, the surrounding gas is gradually pushed away by its increasing gravity. The growing planet can therefore open a partial gas gap and reverse the local gas pressure gradient, halting the inward drift pebbles. The planet itself is hence isolated from pebble accretion. The planet mass initiating for the termination of pebble accretion is defined as the pebble isolation mass \citep{Lambrechts2014b}.
\cite{Bitsch2018} conducted  $3$D hydrodynamic simulation and obtained a fitting formula  
 \begin{equation}
   \begin{split}
 M_{\rm iso} = & 25 \left( \frac{h_{\rm g}}{0.05} \right)^3 \left( \frac{M_{\star}}{\Ms} \right) \left[ 0.34 \left(  \frac{ -3}{ {\rm log} \alphat}  \right)^4 + 0.66 \right] \\
  & \left[ 1-   \frac{ \partial  {\rm ln } P /  \partial {\rm ln} r +2.5} {6}        \right] \Me.
 \label{eq:m_iso}
 \end{split}
 \end{equation}
 When neglecting the pressure gradient dependence and adopting $\alpha_{\rm t} =10^{-3}$, we rewrite the pebble isolation mass as  
 \begin{equation}
  M_{\rm iso}  = 25 \left( \frac{h_{\rm g}}{0.05} \right)^3 \left( \frac{M_{\star}}{\Ms} \right).
   \label{eq:m_iso_sim}
  \end{equation}
We note that  $  M_{\rm iso} \propto M_{\star}^{1+ \beta}$, where $h_{\rm g} \propto M_{\star}^{\beta/3}$. Although  $h_{\rm g} $ might weakly correlate with $M_{\star}$ (depends on detailed disk structure),  to the zero order approximation,  the pebble isolation mass  linearly scales with the stellar mass.   

From \eqs{h_vis}{h_irr} we find that in the outer stellar irradiated disk region $M_{\rm iso}$ is independent of time and increases with the distance  ($M_{\rm iso} \propto r^{6/7}$), while in the inner viscously heated disk region the isolation mass decreases with time and is weakly dependent on the distance  ($M_{\rm iso} \propto r^{-3/16}$).

\subsubsection{Gas accretion}
\label{sec:gas}
We adopt a similar gas accretion prescription as  \cite{Ida2018} and \cite{Johansen2018}. 
The low-mass planet contains a tiny gas atmosphere in a hydrostatic equilibrium.  Once the core mass excesses the critical mass ($\simeq 10-20\Me$), the envelope collapses in a runaway manner. We assume this Kelvin-Helmholtz contraction starts when the core mass reaches $M_{\rm iso}$.
The gas accretion rate is based on the work of \cite{Ikoma2000}, 
 \begin{equation}
  \begin{split}
 \left(\frac{d M_{\rm p, g}}{dt}\right)_{\rm KH} =10^{-5} \left( \frac{\Mp }{10 \Me} \right)^4 \left( \frac{\kappa_{\rm env}}{1 \rm \ cm^2g^{-1}}  \right)^{-1} \rm \  M_{\oplus}\,yr^{-1},
 \label{eq:gas_KH}
 \end{split}
 \end{equation}
 where $\kappa_{\rm env}$ is the opacity in the planet's gas envelope. This value can be significantly lower than that in the protoplanetary disk ($\kappa$), due to the efficient coagulation and sedimentation of dust grains in planetary envelopes  \citep{Ormel2014,Mordasini2014}. We adopt $\kappa_{\rm env} = 0.05 \ \rm cm^{2} g^{-1} $ and assume it does not vary with the disk metallicity. 

In \eq{gas_KH} the accretion rate increases rapidly when the planet mass increases. This situation is only appropriate when there is a sufficient supply of the disk gas. 
However, the gas accretion is regulated by how much disk gas can enter the planet Hill sphere. 
High resolution hydrodynamic simulations found that only fraction of gas within the planet's Hill sphere can be accreted \citep{Tanigawa2002,Machida2010}.  
We propose a simple accretion prescription that 
 \begin{equation}
\left(\frac{d M_{\rm p,  g}}{dt}\right)_{\rm Hill} =f_{\rm acc} v_{\rm H} R_{\rm H} \Sigma_{\rm Hill} = \frac{f_{\rm acc}}{3 \pi} \left( \frac{R_{\rm H}}{H_{\rm g}} \right)^2 \frac{\dot M_{\rm g}}{\alphag} \frac{\Sigma_{\rm gap}}{\Sigma_{\rm g}},
\label{eq:gas_hill} 
\end{equation}
 where $v_{\rm H} = R_{\rm H} \Omega_{\rm K}$ is the Hill velocity and $R_{\rm H} = (M_{\rm p}/3M_{\star})^{1/3} r$ is the Hill radius of the planet, $\Sigma_{\rm Hill}$ is the gas surface density nearly the planet Hill sphere.  In this paper, the fraction of gas in Hill sphere that can be accreted is parameterised by $f_{\rm acc}$, which is set to be $0.5$. It is worth pointing out that our formula differs from \cite{Ida2018}, where they adopted a $(R_{\rm H}/H_{\rm g})^{2}$ dependence on $f_{\rm acc}$ (their Eq.  21). The disk surface density in the vicinity of the planet $\Sigma_{\rm Hill}$  is equal to $\Sigma_{\rm g}$ for the low-mass planet that does not modify the local gas density, while it becomes  $\Sigma_{\rm gap}$ for  the massive planet that opens a gap and depletes the surrounding gas. We explain  the planet mass dependence on  $\Sigma_{\rm gap}/\Sigma_{\rm g}$ in \se{type_II}.

Furthermore, the gas accretion is limited by the total gas flux in the disk that crosses the orbit of the planet.  
 To summarize,  the gas accreted onto the planet is expressed as  
  \begin{equation}
 \dot M_{\rm p, g} = \min \left[ \left(\frac{d M_{\rm p, g}}{dt}\right)_{\rm KH} ,\left(\frac{d M_{\rm p, g}}{dt}\right)_{\rm Hill}, \dot M_{\rm g}    \right].
 \label{eq:gas_acc}
 \end{equation}  
 The Kelvin-Helmholtz contraction is initially dominated for the low-mass planet. When the planet grows and $(d M_{\rm p g}/ dt)_{\rm KH}$ is larger than  $(d M_{\rm p g}/dt)_{\rm Hill}$, the gas accretion is limited by the amount of  gas within the planet's Hill sphere. Once the planet  grows sufficiently massive, the available gas is eventually determined by the global accreting flow across the orbit of the planet. 
 
 \subsection{Planet migration}
\label{sec:migration}
Planets in protoplanetary disks interact gravitationally with the disk gas and undergo orbital migration (see \cite{Kley2012,Baruteau2014} for reviews).
We describe the type I migration for  low-mass planets in \se{type_I}, the gap opening criterion and  the type II migration for massive planets in  \se{type_II}, respectively.

\begin{figure}[t]
    \includegraphics[scale=0.48, angle=0]{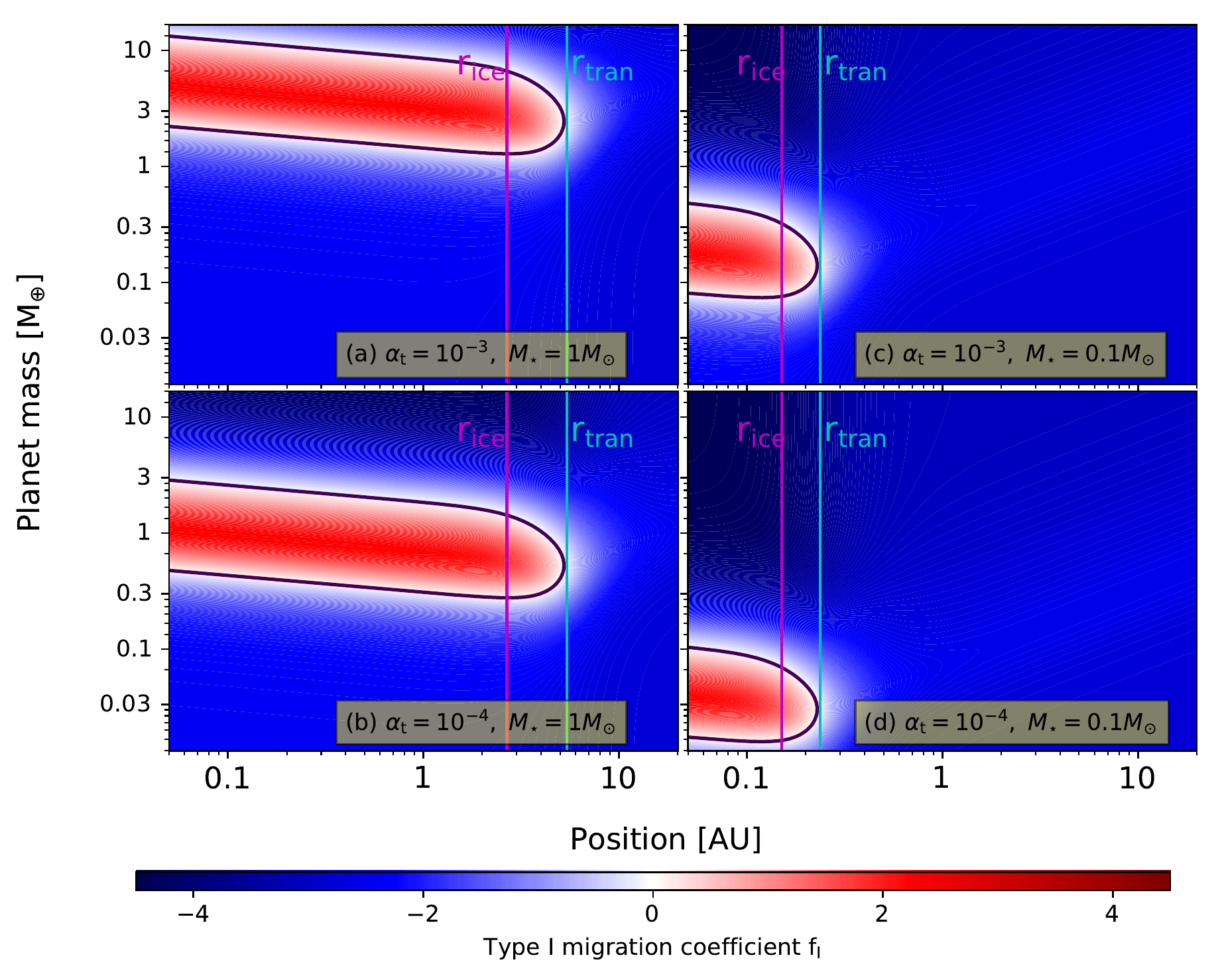}
	\caption{   Type I migration coefficient $f_{\rm I}$ as functions of the planet mass and disk radius. The red (blue) means the migration is outward (inward).  
	The left and right columns show typical disk models with  $\dot M_{\rm g} = 3 \times 10^{-8} \Msyr$  around $M_{\star} = 1\, M_{\odot}$ stars, and  $\dot M_{\rm g} = 3 \times 10^{-10} \Msyr$ around a $M_{\star} = 0.1\, M_{\odot}$,  respectively. The top and bottom rows are disk turbulent viscosity $\alphat$ of $10^{-3}$ and $10^{-4}$. The black line refers to the zero-torque location, the magenta and cyan lines are $r_{\rm ice}$ and $r_{\rm tran}$, respectively.
	 Other parameters are $\alphag =10^{-2}$, $\kappa_0 = 10^{-2}$ and  $L_{\star} -M_{\star}^2$.	 The effect of outward migration is weaker for planets in less turbulent disks and/or around less massive stars. 
	}
	\label{fig:typeI}
\end{figure}

\subsubsection{Type I migration}
\label{sec:type_I}
The type I migration rate can be expressed as 
  \begin{equation}
\dot r_{\rm I} = f_{\rm I} \left( \frac{\Mp}{M_{\star}} \right)  \left( \frac{\Sigma_{\rm g} r^2}{ M_{\star}}  \right) h_{\rm g}^{-2} v_{\rm K},
\label{eq:type_I}
\end{equation}  
where   $f_{\rm I} (s,p, \alphat)$ is the migration prefactor adopted from \cite{Paardekooper2011}. In their findings the diffusion of the gas across the planet horseshoe region takes considerable role on angular momentum transportation and could lead to outward migration. The strength and direction of migration is determined by a combined of  gas surface density and temperature gradients ($s$ and $p$) as well as  the local turbulent viscosity parameter ($\alphat$). In the inner viscously heated region, the planet can migrate outward \citep{Kretke2012,Liu2015} to the transition radius $r_{\rm tran}$ when it approaches an optimal mass  
\begin{equation}
   \begin{split}
M_{\rm opt} = 5
\left(\frac{\alphat} {10^{-3}} \right)^{2/3}
\left(\frac {h_{\rm g}}{0.05} \right)^{7/3}
\left(\frac{M_{\star}}{1 \ M_{\odot}} \right) 
\ M_{\oplus}.
\label{eq:m_opt}
   \end{split}
\end{equation}
Based on \eqs{rtrans}{m_opt}, the optimal mass  for outward migration at $r_{\rm tran}$ can be expressed as
\begin{equation} 
  \begin{split}
M_{\rm opt,tran} \simeq  & 2
 \left( \frac{\dot M_{\rm g}}{10^{-8} \Msyr} \right)^{0.48} 
\left(\frac{M_{\star}}{1 \ M_{\odot}} \right)^{0.16} 
\left(\frac{L_{\star}}{1 \ L_{\odot}} \right)^{0.07}\\
& \left(\frac{\alphag} {10^{-2}} \right)^{-0.24} 
 \left(\frac{\kappa} {10^{-2}} \right)^{0.24}
\left(\frac{\alphat} {10^{-3}} \right)^{0.67}
\ M_{\oplus}.
\label{eq:m_opt_rtran}
 \end{split}
\end{equation}

 \Fg{typeI} illustrates the type I migration map for systems around different stellar masses and turbulent viscosities.  In \fg{typeI}, $M_{\rm opt}$ is higher and  $r_{\rm tran}$, $r_{\rm ice}$ are further out in disks around solar-mass stars than those around very low-mass M-dwarfs. Interior to $r_{\rm tran}$ planets with $M_{\rm p}$ $\sim $ $M_{\rm opt}$ undergo outward migration while planets beyond  $r_{\rm tran}$ migrate directly inward. Thus, planets growing to $M_{\rm opt}$ would migrate to and temporarily stay at $r_{\rm tran}$ in the two-component disk. They have a longer duration to accrete materials before entering  into the inner disk cavity. As shown in \fg{typeI} this outward migration effect is more significant in higher turbulent disks and/or around more massive stars.

\subsubsection{Gap-opening  and type II migration}
\label{sec:type_II}
When the planet becomes massive enough, it strongly perturbs the surrounding gas and opens an annular gap. Recent hydrodynamic simulations explored the  dependence of the depth of the gap on the planet mass, the disk aspect ratio and the viscosity \citep{Duffell2013,Fung2014,Fung2016}.  \cite{Kanagawa2015} derive the analytical formula of the gap depth created by a planet is that 
\begin{equation}
 \frac{\Sigma_{\rm g}}{\Sigma_{\rm gap}} =1 + 0.04 \left( \frac{M_{\rm p}}{ M_{\star}}\right)^{2} \left(\frac{1}{h_{\rm g}} \right)^{5}  \left(\frac{ 1}{\alphat} \right),
\label{eq:sig_gap}
\end{equation}
where $\Sigma_{\rm g}$ and $\Sigma_{\rm gap}$ are the unperturbed gas surface density and the surface density at the bottom of the gap.

 The gap-opening mass is defined as when the gas density at the bottom of the gap is reduced to $50\%$ of the unperturbed value, $\Sigma_{\rm gap}/ \Sigma_{\rm g} =0.5$.  Followed by \cite{Kanagawa2015}, the gap-opening mass is expressed as
\begin{equation}
M_{\rm gap} =  30 \left(  \frac{\alphat}{10^{-3}} \right)^{1/2} \left( \frac{h_{\rm g}}{0.05} \right)^{5/2} \left( \frac{M_{\star}}{\Ms}  \right) \Me.
\label{eq:m_gap}
\end{equation}
Noteworthy, the planet with a pebble isolation mass  opens a shallow gap and truncates the drifting pebbles,  which roughly corresponding to a gap depth of  $\approx 15 \%$ \citep{Bitsch2018}. Principally, $M_{\rm iso}$ should be lower than $ M_{\rm gap}$. \cite{Johansen2018} simulated a planet  embedded in a $1$D accretion disk with torque formulas adopted from \cite{D'Angelo2010}. They found that the $1$D gap opening mass is $2.3$ times of the pebble isolation mass, where these two quantities are defined as the relative gap depth is $50\%$ and $15 \%$, respectively (see their Fig. 3).

However,  the $\alphat$-dependence of the pebble isolation mass from  $3$D hydrodynamic of  \cite{Bitsch2018} differs  from  
  \cite{Kanagawa2015}'s gap-opening mass, in particular at low $\alphat$. When compared \eq{m_iso} with \eq{m_gap}, $M_{\rm iso}$ is even lower than $M_{\rm gap}$ at $\alphat=10^{-4}$. The shallower $\alphat$-dependence on \cite{Bitsch2018}'s isolation mass in less turbulent cases is probably related to Rossby wave instabilities and vortex formation at the gap edge \citep{Hallam2017}. To be consistent with \cite{Johansen2018}, we adopt the pebble isolation mass from \cite{Bitsch2018} and set $M_{\rm gap}= 2.3 M_{\rm iso}$. Thus, both $M_{\rm gap}$ and $M_{\rm iso}$ decreases  by  $25 \%$ when $\alphat$ drops from $10^{-3}$ to $10^{-4}$.

In the classical picture type II migration \citep{Lin1986}, the gas is thought to be unable to across the gap. The planet is therefore locked to the viscous evolution of the disk gas. However, some hydrodynamic simulations found that even for a deep gap opened by a Jupiter-mass planet, the surrounding gas could still pass through the gap \citep{Duffell2014,Durmann2015}.  As a result, the migration of the planet in some circumstances can even be faster than the radial drift of the disk gas \citep{Robert2018}.       
    
Conducting hydrodynamical simulations with broad ranges of disk parameters and different planet masses, \cite{Kanagawa2018} proposed a new physical picture of type II migration. The  torque for a gap-opening planet can be expressed as $ \Gamma = \Gamma_0  (\Sigma_{\rm gap}/ \Sigma_{\rm g}) $
where the nominal type I torque reads $\Gamma_0= (M_{\rm p}/M_{\star})^{2} h_{\rm g}^{-2} \Sigma_{\rm g} a_{\rm p}^4 \Omega_{\rm K}^2$.
This implies that the torque exerted on the planet depends on the gas density at the bottom of the gap. Thus, the corresponding migration for a very massive planet slows down as $\Sigma_{\rm gap}$ decreases, and the rate can be slower compared to the classical type II migration.

Furthermore, one uniform migration prescription can be given for both type I ( the low-mass planets) and type II (the gap-opening massive planets) regimes.  The migration rate of the planet can be written as
\begin{equation}
\dot r =  \dot r_{\rm I}/ \left[ 1 + \left(\frac{M_{\rm p}}{M_{\rm gap}}\right)^2 \right].
\label{sec:migration rate}
\end{equation}
From the above expression,  $ \dot r_{\rm }  \propto M_{\rm p} $ in the type I regime whereas  $\dot r_{\rm} \propto M_{\rm p}^{-1} $ in the type II regime.

The inner disk is truncated by stellar magnetospheric disk interaction \citep{Lin1996}. 
We assume a static inner disk edge (magnetospheric cavity) that approximates to the co-rotation radius of the central star  \citep{Mulders2015}, 
\begin{equation}
r_{\rm in} = \sqrt[3] {\frac{ G M_{\star}} {\Omega_{\star}^2} }= 0.04 \left (\frac{M_{\star}}{ \Ms}\right)^{1/3} \AU.
\label{sec:migration rate}
\end{equation}
where $\Omega_{\star}$ is the stellar spin frequency, and the rotation periods of the young T Tauri stars range from  $1$ to $10$ days \citep{Herbst2002}.
 We assume that when planets enter $r_{\rm in}$, their migration and mass growth are quenched. We also discuss the situation when low-mass planets stall at the outer edge of the magnetospheric cavity \citep{Liu2017,Romanova2019} and keep  accreting pebbles in \ap{migration}.

\section{Illustrative simulations}
\label{sec:result}

\begin{table}
    \centering
    \caption{Simulations set-up in \se{result}}
    \begin{tabular}{lclclclclcl}
        \hline
        \hline
        Name    &  $M_{\star} $           & $R_{\rm d0} $  & $\dot M_{\rm g0}$   & $\xi$   \\ 
         &        ($ \rm M_{\odot}$)    &  (AU)   & ($ \rm M_{\odot} \, yr^{-1}$)  & & &    \\
      \hline
        \texttt{run\_R1} &  1.0 & 30  & $ 6 \times 10^{-8}$  & 0.01  \\
        \texttt{run\_R2} &  1.0 & 100  & $ 6 \times 10^{-8}$  & 0.01    \\
        \texttt{run\_R3} &  1.0 & 160  & $ 6 \times 10^{-8}$  & 0.01   \\
        \texttt{run\_D1} &  1.0 & 100  & $2 \times 10^{-8}$  & 0.01   \\
        \texttt{run\_D2} &  1.0 & 100 & $5 \times 10^{-8}$  & 0.01   \\
        \texttt{run\_D3} &  1.0 & 100 & $ 10^{-7}$  & 0.01    \\
        \texttt{run\_Z1} &  1.0 & 80  & $5 \times 10^{-8}$  & 0.003  \\
        \texttt{run\_Z2} &  1.0 & 80  & $5 \times 10^{-8}$  & 0.01    \\
        \texttt{run\_Z3} &  1.0 & 80  & $5 \times 10^{-8}$  & 0.03   \\
         \texttt{run\_M1} &  1.0 & 100  & $10^{-7}$  & 0.03  \\
        \texttt{run\_M2} &  0.3 & 100  & $1.1 \times 10^{-8}$  & 0.03    \\
        \texttt{run\_M3} &  0.1 & 100  & $1.6 \times 10^{-9}$  & 0.03    \\
          \hline
        \hline
    \end{tabular}
    \label{tab:illustration}
\end{table}

\begin{figure}[t]
      \includegraphics[scale=0.43, angle=0]{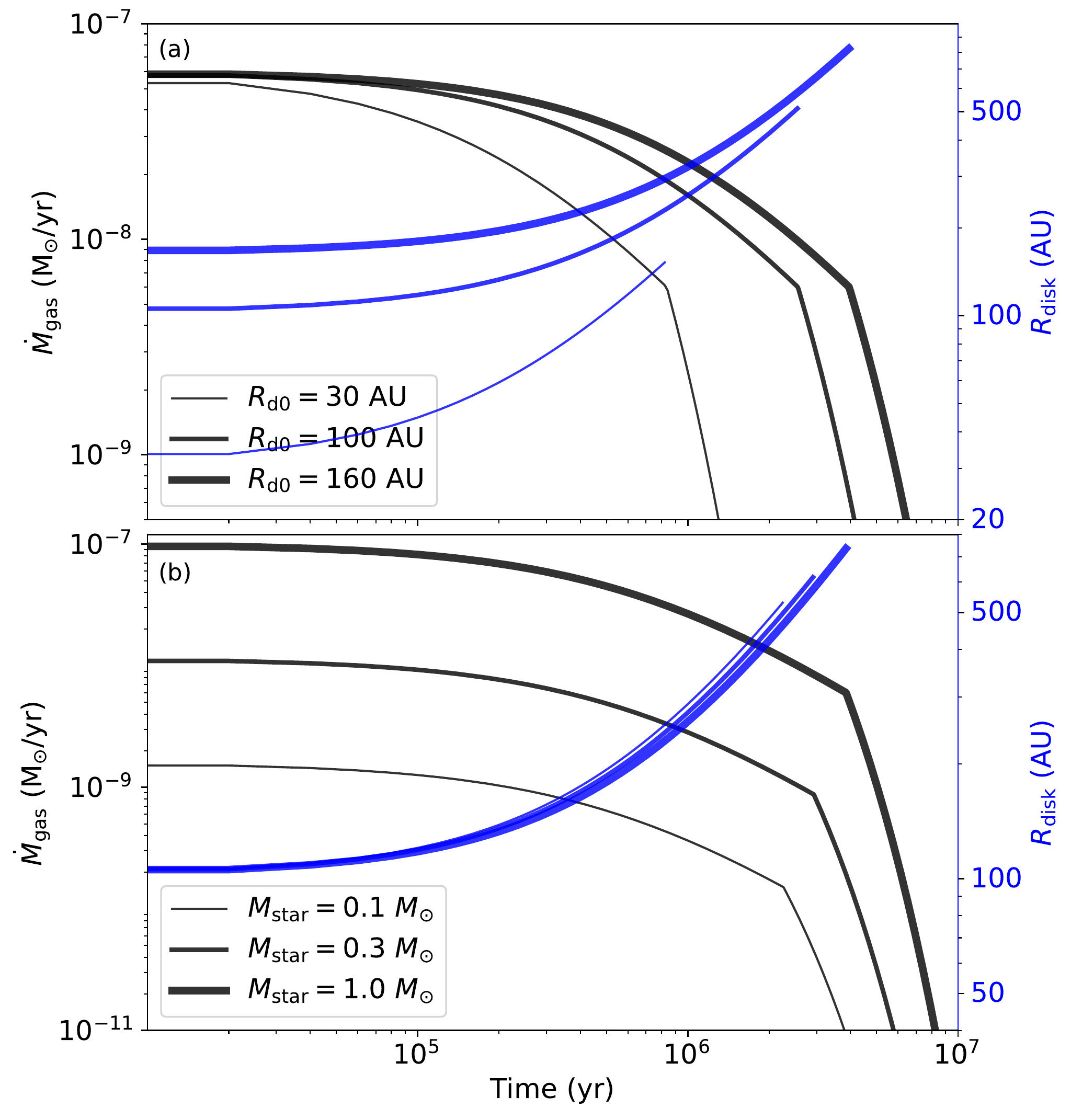}
           \caption{
           Evolution of disk accretion rate (black) and characteristic disk size (blue). When $\dot M_{\rm g} < \dot M_{\rm pho} $ (\eq{pho}), the disk angular momentum is transported by viscous accretion; later the gas removal is driven by stellar X-ray evaporation. The expansion of the disk sizes are shown in solid lines due to viscous spreading.  
      Top: three systems with different initial disk sizes ($R_{\rm d0}= 30$ AU, $100$ AU and $160$ AU) are shown around a $1 \Ms$ star. The initial disk accretion rates are all  $6 \times 10^{-8} \Msyr$. 
      Bottom: three systems with the same initial disk size ($R_{\rm d0}= 100$ AU) are shown around different masses of stars ($M_{\star}= 1 \ M_{\odot}$, $0.3 \ M_{\odot}$ and $0.1 \ M_{\odot}$). The initial disk accretion rates follow $10^{-7} (M_{\star}/M_{\odot})^{1.8}  \Msyr$.  Note that the two panels share the same x-axis.}
\label{fig:disk}
\end{figure}

In this section we focus on the growth and migration of a single protoplanetary embryo,  which is born at the water ice line for different disk conditions and around different masses of stars. The influence of four key model parameters are investigated subsequently: the initial characteristic disk size $R_{\rm d0}$ in \se{size}, the initial disk accretion rate $\dot M_{\rm g0}$ in \se{mdot}, the disk metallicity $Z_{\rm d}$ in \se{metallicity}, and the stellar mass $M_{\star}$ in \se{mass}.

 Since we will study the planet formation around different types of stars,  the stellar mass dependence on the protoplanetary disk properties, such as the gas disk accretion rate $\dot M_{\rm g}$, the stellar  luminosity $ L_{\rm s}$ and the initial disk size $R_{\rm d0}$ are described as follows. 
  \begin{enumerate}[1.]
  \item $\dot M_{\rm g}-M_{\star}$:   We assume $\dot M_{\rm g} \propto M_{\star}^{1.8}$. This empirical correlation is obtained from the measurements of gas accretion onto young stars in different star formation regions, based on stellar UV flux excess and spectroscopic method \citep{Hartmann1998,Muzerolle2003,GarciaLopez2006,Natta2006,Alcala2014,Manara2016}.    There is however no evidence that indicates any $\dot M_{\rm g} - Z_{\star}$ correlation. 
  
 \item $ R_{\rm d0}-M_{\star}$: One way to probe the disk size is from the dust continuum emissions. However, even the dust particles are well coupled and mixed with gas at the beginning, as particles grow and drift they gradually decouple from the gas, resulting in a smaller dust disk compared to the gas disk.  The other way is to measure the disk size from CO gas line emissions. Recently, \cite{Ansdell2018} used the samples of young Lupus star-forming region and found that the gas disk radii appear to be independent of the masses of their host stars.  Conducting radiation hydrodynamical simulations of star cluster formation, \cite{Bate2018} also found that the the sizes of the early protostellar disks are barely (at most very weakly) dependent on $M_{\star}$. For this work we assume that $R_{\rm d0}$ is independent of $ M_{\star}$.  Combined with the above $\dot M_{\rm g}-M_{\star}$ correlation, the disk masses are inferred to be superlinearly correlated with  the masses of their stellar hosts.  This is also supported by the ALMA disk mass measurements from dust continuum emissions \citep{Pascucci2016,Manara2016}.    
  
     \item  $L_{\star} -M_{\star}^{p}$:   The intrinsic luminosities of embedded young stars are hard to probe. Large observational uncertainties arise from the pollution of infall materials and high extinction from the surrounding dust.  Theoretical models predict a time evolving  $L_{\star}-M_{\star}$  correlation \citep{Burrows1997,Baraffe1998}. For the stars with ages less than a few Myr, $L_{\star}$  roughly increases with $M_{\star}^{1-2}$ (see Fig.2 of \cite{Baraffe2002}). When stars gradually cool and enter the main-sequences, $L_{\star}$ transitions to a dependence  $M_{\star}^{3-4}$. Since the time span of our simulation is typically shorter than the stellar pre-main-sequence phase, for simplicity,  we adopt $p= 2$  for the fiducial model. The circumstance of  linear $L_{\star}-M_{\star}$ correlation ( $p=1$) is also examined in \ap{luminosity} .
   \end{enumerate}

\begin{figure*}[t]
        \includegraphics[scale=0.55, angle=0]{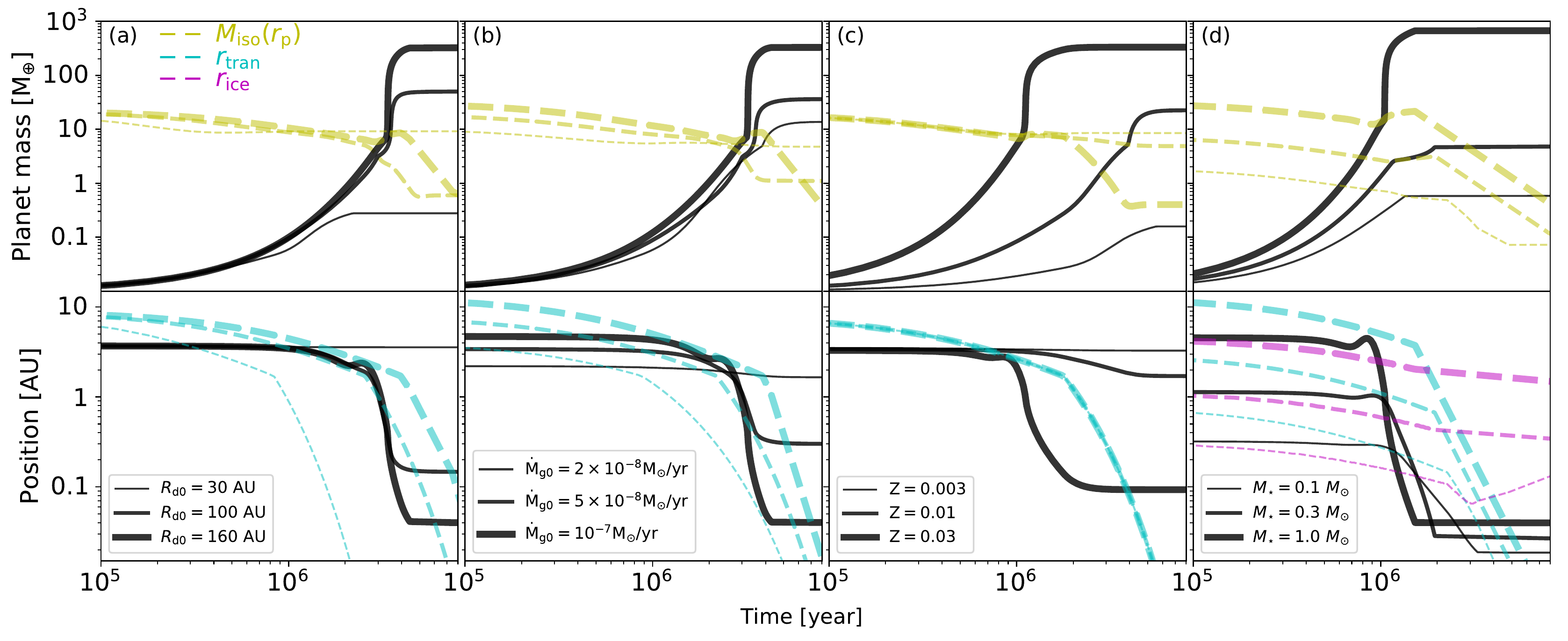}
           \caption{
        Planet mass growth (top) and orbital  evolution (bottom)  with different initial disk sizes (panel a, \texttt{run\_R1} to \texttt{run\_R3}), initial disk accretion rates (panel b, \texttt{run\_D1} to \texttt{run\_D3}), disk metallicities (panel c, \texttt{run\_Z1} to \texttt{run\_Z3}) and masses of central stars (panel d, \texttt{run\_M1} to \texttt{run\_M3}).
        The yellow dashed lines represent the pebble isolation masses at the planets' locations, whereas the cyan and the magenta dashed lines correspond to the transition radii between two disk regions and the water ice lines (only shown in panel d). Massive planets are likely to form when the disk sizes are larger,  the disk accretion rates are higher, the disks are more metal rich, and their stellar hosts are more massive. }
\label{fig:growth}
\end{figure*}

\subsection{Primordial disk size}
\label{sec:size}

 Three disks with different characteristic sizes around solar-mass stars are shown in \fg{disk}a, from $R_{\rm d0}= 30$ AU, $100$ AU to $160$ AU.  The initial gas disk accretion rates are $\dot M_{\rm g0} =  6 \times 10^{-8} \Msyr$.  In the early phase when $\dot M_{\rm g} >  \dot M_{\rm pho} (=6\times 10^{-9} \Msyr)$, the disk accretion rate (black) decreases slowly while the disk size (blue) expands through viscous spreading. When $\dot M_{\rm g} < \dot M_{\rm pho}$, the significant decline of the accretion rate is driven by the stellar photoevaporation.
 
For a small disk of $R_{\rm d0} = 30$ AU,  the accretion rate diminishes by two orders of magnitude within $1.2$ Myr and the disk expands to $120$ AU. For a comparison, the disks expand to $500$ AU and $800$ AU when $R_{\rm d0} = 100 $  AU and  $160$ AU, respectively. The disk gas can last for $4$ Myr and  $6$ Myr  in the latter two cases.  The disk aspect ratio at the characteristic radius $R_{\rm d0}$ is larger for a larger disk, and  from \eq{tvis} the viscous evolution ($t_{\rm vis}$) is  longer. In such a disk, the accretion rate therefore drops more slowly, resulting in a longer disk lifetime.

We simulate the evolution of the single protoplanetary embryo in \fg{growth}, with masses in upper panels and semi-major axes in lower panels. The yellow and green dashed lines represent the pebble isolation mass  at the planet's location $M_{\rm iso}(r_{\rm p})$  and the disk transition radius $r_{\rm tran}$, respectively.  
 
The runs with different initial characteristic disk sizes are shown in \fg{growth}a (\texttt{run\_R1} to \texttt{run\_R3} in \tb{illustration}). 
In the disk of $R_{\rm d0} =30 $ AU, the embryo grows \textit{in-situ} to a mass of  $0.3 \Me$. This is because the disk gas is significantly depleted within $1$ Myr. The planet neither accretes large amounts of pebbles nor migrates substantially within this time. 
The planet in the disk of $R_0 =100 $ AU grows faster. It migrates and remains at the transition radius (green line) until the mass exceeds $\simeq 5 \Me$ (also see  \fg{typeI}a).
After that it migrates inward due to the saturation of the corotation torque \citep{Paardekooper2011}.  The rapid gas accretion  starts when the embryo reaches the pebble isolation mass ($\simeq$$10 \Me$). However, at that time the disk mass is already largely depleted. The planet grows up to $ 50 \Me$ at $t = 3.5$ Myr.
In the case of the disk with $R_0 =160 $ AU, the disk accretion rate remains high for a longer time. In contrast to the $R_0 =100 $ AU case,  the rapid gas accretion can proceed longer when the planet approaches the isolation mass. After $t = 4$ Myr the planet migrates into the cavity and grows into a hot Jupiter.

The disk accretion rate declines more slowly for a larger disk. The pebble flux in such a disk drops slowly for a fixed pebble-to-gas flux ratio.  Therefore, the core growth is more rapidly for a larger disk.  A higher disk accretion rate also indicates a larger disk scale height.  As a result, first, the optimal mass $M_{\rm opt}$ for keeping the planet  at $r_{\rm tran}$ is higher (\eq{m_opt}). This means the planet has a longer duration to accrete pebbles before entering the inner disk cavity. Second, the pebble isolation mass $M_{\rm iso}$ also becomes higher (\eq{m_iso}). This indicates that pebble accretion is quenched for a more massive planet. By then these planets can accrete gas more rapidly (\eqs{gas_KH}{gas_hill}).  Conclusively, a massive planet is easy to form in a larger disk.

\subsection{Disk accretion rate}
\label{sec:mdot}

 Three runs (\texttt{run\_D1} to \texttt{run\_D3}) with different initial disk accretion rates are illustrated in \fg{growth}b. 
 Similar to the effect of $R_{\rm d0}$, when $\dot M_{\rm g0}$ is higher, the embryo grows to a more massive planet. 
In disks with  $\dot M_{\rm g0} = 2\times 10^{-8} \Msyr$  and $ 5 \times 10^{-8} \Msyr$  AU, the planets grow to $13\Me$ and $35 \Me$, respectively.  In the disk with  a higher  $\dot M_{\rm g0}=   10^{-7} \Msyr$, the planet becomes a hot-Jupiter with $M_{\rm p } \simeq 300 \Me$. 

As explained in \se{size},  when the disk accretion rate is higher, the planet grows faster by pebble accretion. In this circumstance $r_{\rm ice}$ and $r_{\rm tran}$ are also further out. The corresponding  $M_{\rm opt}$ and $M_{\rm iso}$ at these locations are higher. Thus,  the embryo migrates outward to a further $r_{\rm tran}$, and remains there until its mass excess a higher $M_{\rm opt}$. The time span for the growth becomes longer.  When the core reaches a higher $M_{\rm iso}$, the gas accretion is also more rapidly. In conclusion, a massive planet is likely to form in a disk with a high $\dot M_{\rm g0}$.  

\subsection{Disk metallicity}
\label{sec:metallicity}

 The disk metallicity has two effects.  First, from \eq{sig_ratio} the disk metallicity scales with the pebble-to-gas flux ratio. For disks with fixed gas accretion rates, a higher $Z_{\rm d}$ indicates a larger mass flux of pebbles.  Second, the disk opacity also correlates with the grain abundance.
 In order to separate different physical effects, we only adopt the first condition ($ Z_{\rm d} = \xi$)  and remain the disk opacity unchanged here. The metallicity effect on $\kappa$ will be considered in \se{obs}.
 
The growth of the embryos with three different disk metallicities  ($Z = 0.003$, $0.01$ and $0.03$) are illustrated in \fg{growth}c (\texttt{run\_Z1} to  \texttt{run\_Z3}).
The effect of  the disk metallicity is clear.   A factor of ten increase in the metallicity leads to more than three orders of magnitude growth on the planet mass. 
The pebble mass flux is higher in a metal rich disk ($\dot M_{\rm peb} = \xi \dot M_{\rm g}$).  In this circumstance the core mass grows faster by pebble accretion and therefore attains the isolation mass at more early stage. Since $M_{\rm iso}$ is higher in the early phase when the disk is hotter ($h_{\rm g}$ is larger), the planet in the metal rich disk attains a larger core mass and thus enable to accrete more surrounding gas.  The formation of  a massive planet  is efficient in the metal rich disk.

\subsection{Stellar mass}
\label{sec:mass}

 Disks around  stars of $M_{\star}= 1 \ M_{\odot}$, $0.3 \ M_{\odot}$ and  $0.1 \ M_{\odot}$  are shown in \fg{disk}b.  The initial characteristic sizes are all $100$ AU, and the initial disk accretion rates are $\dot M_{\rm g0} = 10^{-7} (M_{\star}/M_{\odot})^{1.8} \Msyr$.  Based on \eqs{h_irr}{tvis}, $t_{\rm vis}$ is barely dependent on stellar mass.  The viscous evolution and size expansion therefore exhibit very similar trends for the above three cases in  \fg{disk}b.  The disk gas is depleted more severely around more massive stars during the late stellar photoevaporation phase (\eq{pho}).

 The growth of the planets around the above three stars is shown in \fg{growth}d (\texttt{run\_M1} to  \texttt{run\_M3}). Due to the inward migration of $r_{
 \rm ice}$ and planet outward migration, all planets are outside of  $r_{\rm ice}$ (magenta line) in the early stage.  They finally migrate into the inner disk cavity, and end up with $0.7 \Me$, $6 \Me$, and  $800 \Me$ for the host stars of $0.1 \Ms$, $0.3 \Ms$ and $1 \Ms$, respectively.  
   
 Embryos grow more slowly and become less massive around low-mass stars.  This is because the disk around a less massive star has a lower gas accretion rate. The pebble flux is lower and therefore the planet accretes pebbles at a slower rate. Importantly, since $M_{\rm iso} \propto M_{\star}$, the pebble isolation mass is also lower in the system with a less massive central star. For the illustrated systems around stars with $M_{\star} =0.1 \ M_{\odot} $ and  $0.3 \ M_{\odot} $,  $M_{\rm iso} $ is smaller than $10\Me$ and this core mass is not massive enough to trigger runaway gas accretion. Therefore, these planets only accrete modest atmosphere. On the contrary, the pebble isolation mass reaches  $\simeq 20 \Me$ for a solar-mass star.  In this circumstance the planet can initiate the rapid gas accretion and form a massive gas giant planet. 

To conclude,  only  planets with $M_{\rm iso} \gtrsim  10 \Me$ have the ability to efficiently accrete gas and become gas giants.  This condition is more easy to satisfy for systems around GK and early M stars rather than late M stars.  Importantly, $M_{\rm iso}$ can be used as an upper mass limit  for the super Earth planets with $M_{\rm p} \lesssim 10 \Me$.

\section{Migration and composition map}
\label{sec:map}

In the previous section we have discussed the planet formation for different disk parameters and stellar masses. The embryos are all initially placed at $r_{\rm ice}$ and are started from $t=0$ yr. Here we further investigate  the growth and migration of planets when starting from  different time and disk locations. 
Simulations illustrated for comparison are among the stars of two different masses ($M_{\star} = 1 \Ms$ and $0.1 \Ms$) and the disks of two turbulent viscosities ($\alphat =10^{-3}$ and $10^{-4}$).
The final masses of resulting planets and the corresponding water contents in planetary cores are discussed  in \se{growthmigration} and  \se{water}, respectively. Through the following map analysis, we have a better understanding of which types  and how these planets would form by given conditions.

\subsection{Final planetary mass}
\label{sec:growthmigration}

 We vary the birth time $t_0$ and the position $r_0$ of the protoplanetary embryo.  \Fg{mapmass} shows the planet growth maps for different $r_0$ and $t_0$. Two types of stars  ($1 \Ms$ and $0.1 \Ms$)  and two disk turbulent viscous $\alphat $ ($10^{-3}$ and $10^{-4}$) are illustrated.  Color gives the final planet mass and the black line refers to its final location. This plot exhibits what type of planet ($M_{\rm p}$ and $a_{\rm p}$) the embryo eventually grows into for given $t_0$ and $r_0$.

 First we focus on the planet formation around a solar mass star. In \fg{mapmass}a we find that the embryos can grow into Jupiter-mass giant planets when $r_0 \sim r_{\rm ice}$ and $t_0 \lesssim 1$ Myr (red region).  Embryos only grow moderately up to a few Earth mass when they are born at $r_0 \lesssim1$ AU or  $20 \AU  \lesssim r_0 \lesssim 30 \AU$ (green regions). The embryos however seldom grow and migrate when they form late and at wide orbital distances.   Compared to \fg{mapmass}a, the planet growth is generally more efficient in \fg{mapmass}b when the disk turbulence is ten times lower. In \fg{mapmass}b the  formation zone (parameter space) for the gas giant planet is larger and shifts further out, and the super-Earth formation zone is more extended as well.
 
 The disk turbulent strength $\alphat$ is a key parameter, which affects both the planet core growth and the orbital migration. First, the pebble scale height becomes smaller when the disk turbulence is weaker. In this case, since the pebbles settle towards to a thinner vertical layer, pebble accretion is more efficient in the $2$D regime. The mass growth is thus faster. 
 Second, the optimal planet mass for outward migration correlates with $\alphat$. As shown in Fig. \fgnum{typeI}a and  \fgnum{typeI}b,  planets with $M_{\rm p} \simeq 10 \Me$  migrate inward at $\alphat =10^{-4}$,  while they can migrate outward at $\alphat=10^{-3}$ due to the strong, positive unsaturated corotation torque.  Thus, in order to grow gas giant planets, embryos need to form sufficient further out to avoid rapid  inward migration when the disk turbulence is low.  That is why in \fg{mapmass}b the gas giant planet formation zone moves to the $ 20 \AU-40 \AU$ region.  Although the migration of gas giants becomes slower with an increasing $M_{\rm p}$ \citep{Kanagawa2018},  it is still not strong enough to suppress the inward migration, resulting in many close-in gas giant planets.

Figure\fgnum{mapmass}c and \fgnum{mapmass}d illustrate the growth map of planets around a $0.1\Ms$ star. In this case earth-mass planets can form at early stage ($t_0 \lesssim 2$ Myr) and  close to $r_{\rm ice}$ and $r_{\rm tran}$ ($0.1 \lesssim r_0 \lesssim 1 $ AU).  When forming late ($t_0 \simeq 3$ Myr) at  $r_{\rm ice}$, the embryos only grow into $\sim 0.3 \Me$. The growth fails  when embryos are born beyond $2$ AU.  The earth-mass planet formation zone is slightly larger in \fg{mapmass}d compared to \fg{mapmass}c, but both maps exhibit a similar pattern.  As shown in  Fig. \fgnum{typeI}c and \fgnum{typeI}d,  this is because the optimal mass is very low, and hence, the outward migration is insignificant for planets around a $0.1 \Ms$ star at these two adopted $\alphat$.

 Both  $r_0$ and $t_0$ are important for planet formation.  
When the embryos form very close to their central stars, they migrate inward and quickly enter the inner disk edge before reaching the optimal mass to reverse their migration directions. These seeds therefore fail to grow massive. When the embryos form very far away,  pebble accretion becomes inefficient and the core growth is strongly suppressed. Similarly, embryos forming  late grow less since at that time the pebble flux already drops.  However, we also find that late forming embryos ($2-3$  Myr) still grow to planets of $20-30 \Me$ in \fg{mapmass}a.  Notice that the disk lifetime is longer than $3$ Myr in the above case.  When the stellar X-ray photoevaporion dominates, the disk metallicity is enhanced due to rapid gas removal.  It boosts the pebble accretion at late time.

To conclude, embryos forming early and at moderate distances  are most likely to become massive planets.    
For the typical protoplanetary disk parameters,  Jupiter-mass gas giants can form around solar-mass stars, and Earth mass planets can form around  $0.1 \Ms$ stars.  
 
\begin{figure*}[t]
    \includegraphics[scale=0.8, angle=0]{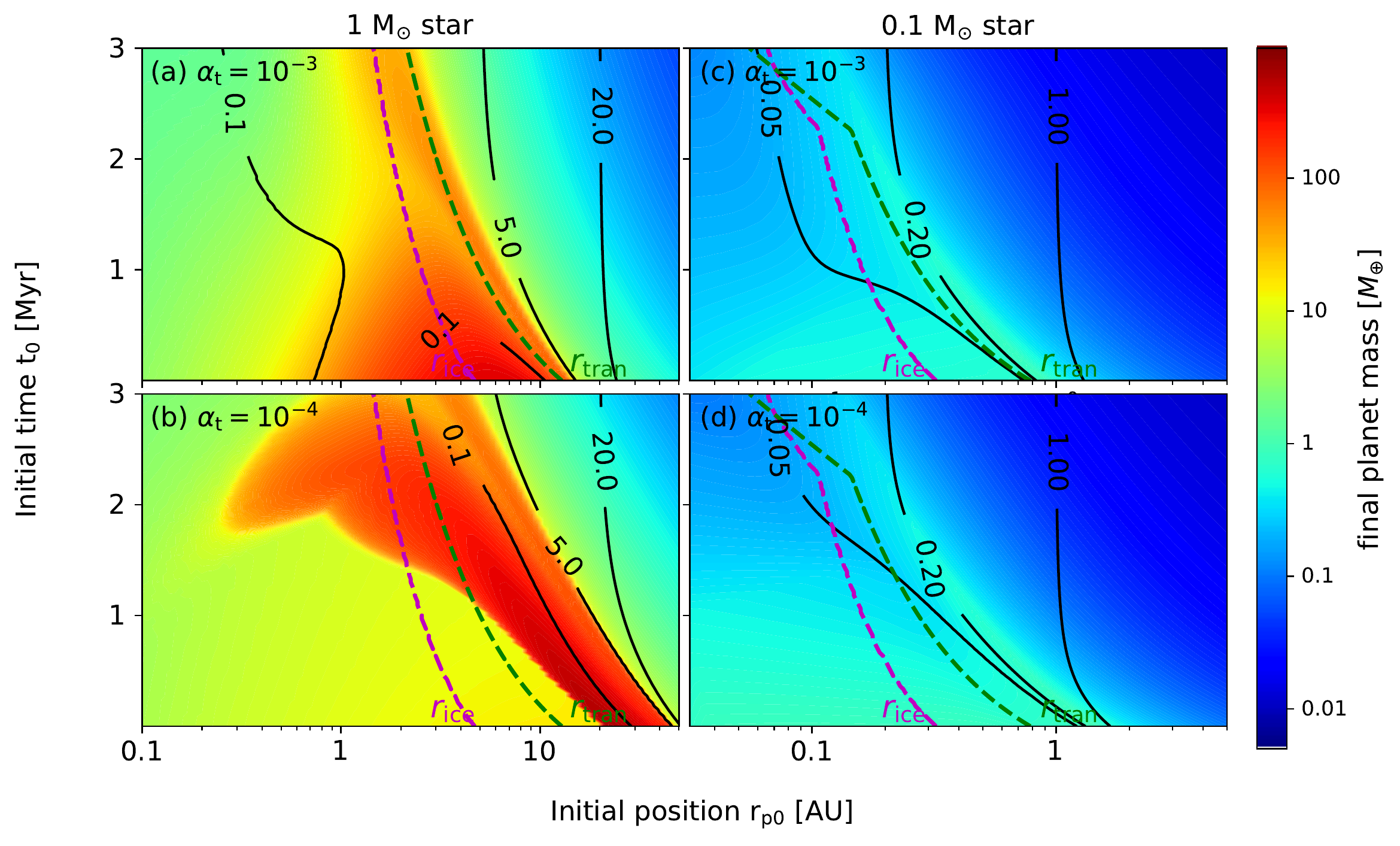}
      \caption{  
      Maps for the growth and migration of planets around stars of  $1 \Ms$ (left) and  $0.1 \Ms$ (right) and at disk turbulent $\alphat =10^{-3}$ (top) and $10^{-4}$ (bottom).
      The  initial time $t_0$ and  initial location $r_0$ are shown in x and y axis. The color corresponds to the final mass of the planet, and the black line represents the final location of the planet.   The water ice line and the transition radius are labeled as magenta and cyan dashed lines. 
       The parameters are adopted from  \texttt{run\_D3} for the left panel and \texttt{run\_M3} for the right panel, respectively. Note that the disk lifetimes in above two cases are longer than $3$ Myr.  Jupiter-mass planets can form around a $1 \Ms$ star while Earth-mass planets can form around a $0.1 \Ms$ star.}
\label{fig:mapmass}
\end{figure*}

 \begin{figure*}[t]
  \includegraphics[scale=0.8, angle=0]{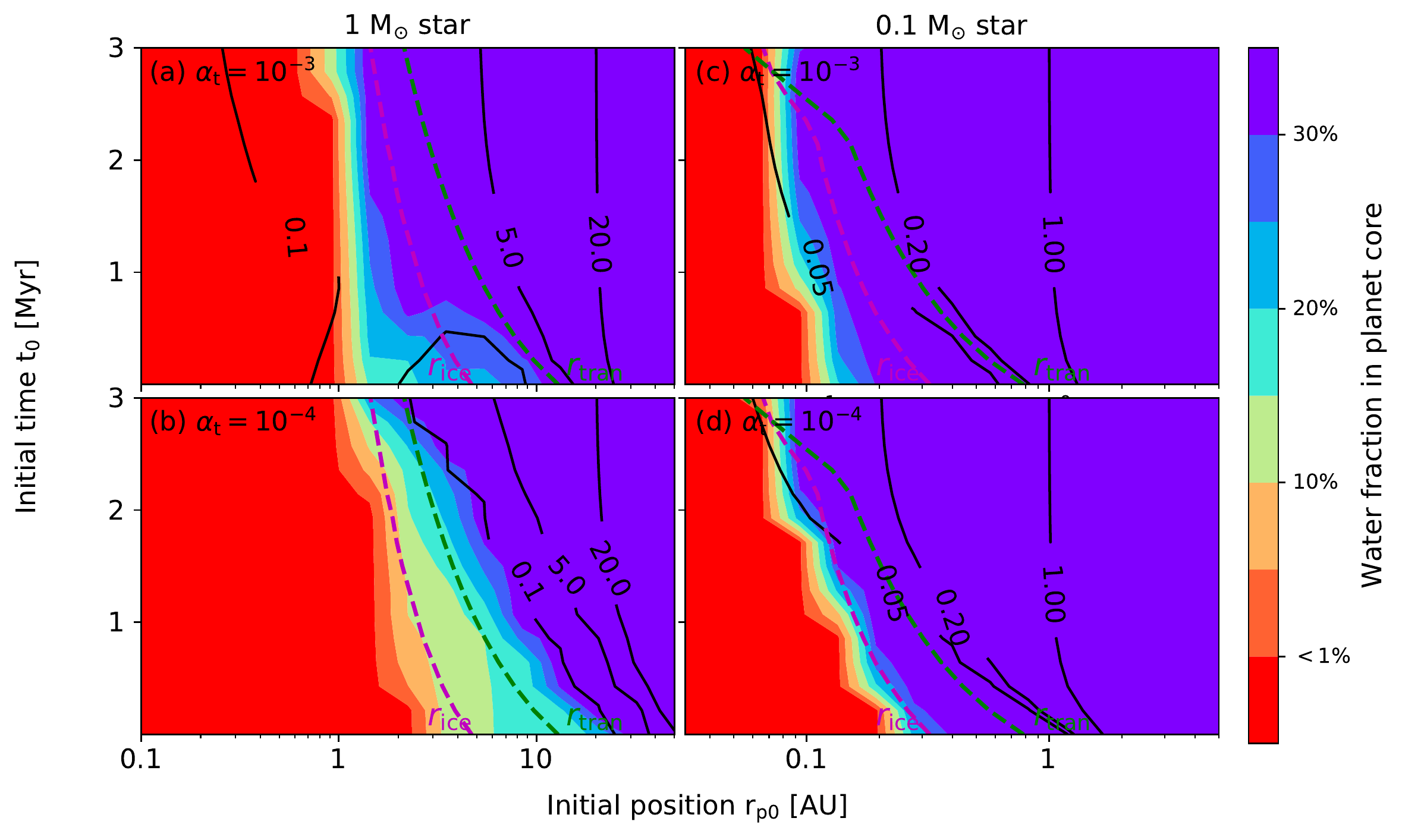}
      \caption{  
      Maps for the core water mass fraction of planets around stars of  $1 \Ms$ (left) and  $0.1 \Ms$ (right) and at disk turbulent $\alphat =10^{-3}$ (top) and $10^{-4}$ (bottom).
      The  initial time $t_0$ and  initial location $r_0$ are shown in x and y axis. The color corresponds to the water fraction, and the black line represents the final location of the planet.   The water ice line and the transition radius are labeled as magenta and cyan dashed lines. 
       The parameters are adopted from  \texttt{run\_D3} for the left panel and \texttt{run\_M3} for the right panel, respectively. The final water of the planets depends on the birth locations of the protoplanetary seeds.  When forming at the water ice line, these planets would end up with a few percent to a few tens of percents  water in mass.  
       }
\label{fig:mapwater}
\end{figure*}

\subsection{Water content}
\label{sec:water}

We assume that the pebbles exterior to the water ice line contain $35\%$ water ice and $65\%$ silicate by mass. When the pebbles drift interior to the water ice line, the ice sublimates and the remaining pebbles are purely rocky. We name these two types "wet" and "dry" pebbles, respectively. The water content in planetary cores depends on where the planets accrete pebbles, and therefore is crucially related with their formation locations as well as the migration history.  

\fg{mapwater} illustrates the water mass fractions in the planetary cores around stars of $1 \Ms$ and $0.1\Ms$ and at $ \alphat=10^{-3}$ and $10^{-4}$, respectively.  We find that embryos initially sufficient close-in  can grow into rocky planets, while embryos born far beyond $r_{\rm ice}$ turn into water-rich planets ($f_{\rm H_2O}\sim 35\%$). 
When embryos form slightly interior or exterior to $r_{\rm ice}$, their water fractions vary in different cases, roughly within $1\%-20\%$ level. The transition between rock-dominated planets ($<1\%$) and water-rich planets ($\gtrsim 10\%$) is quite narrow in terms of $r_0$. 
 Importantly, these ice line planets in less turbulent disks have generally a lower water fraction compared to those in turbulent disks. The difference is more significant for systems around solar-mass stars than low-mass stars. 
 
It is noticed  that the above feature qualitatively agrees with the result shown in Fig. 5  of \cite{Bitsch2019}, despite a different disk model adopted in their work.  Also in their model pebbles have less than $35\%$ of water beyond the ice lines of other volatile species (\eg ,  $\rm NH_3$, $\rm CO_2$),  resulting in a lower water content of distant planets compared to ours.

The resulting feature can be explained as follows. We focus on the radial distance dependence here. Note that $r_{\rm ice}$ decreases with time. This is because the movement of $r_{\rm ice}$ correlates with the disk evolution (\eq{mdot_acc}) on a timescale of $t_{\rm vis}$. For embryos with very close-in initial orbits, their inward migration is always faster than that of $r_{\rm ice}$. These planets are therefore entirely rocky.  However, when embryos form near $r_{\rm ice}$, their growth is slow at the beginning, and the migration of  $r_{\rm ice}$ is faster than these small embryos. They accrete wet pebbles at early phase (also shown in \fg{growth}d).  When these embryos grow massive,  type I migration starts to dominate.   They would undergo outward migration to the exterior of  $r_{\rm ice}$ and accrete wet pebbles for a while. When their masses exceed $ M_{\rm opt}$, they migrate inward rapidly.  They thus accrete dry pebbles after migrating interior of  $r_{\rm ice}$.  Hence, these planets formed near  the ice line finally end up with moderate water fractions. 
For embryos formed at very wide orbits, the migration timescale is too long and they never bypass the ice line, resulting in water-rich cores.

Now we explain the stellar mass and disk turbulence dependence. The optimal mass for outward migration correlates with $\alphat$. Therefore, in \fg{mapwater}a the ice line planets accrete more wet pebbles outside of $r_{\rm ice}$ and become more water-rich compared to those in \fg{mapwater}b. On the other hand,  this outward migration effect is less significant for systems around very low mass stars compared to solar-mass stellar hosts (\fg{typeI}). The water fraction of the planets around $0.1 \Ms$ stars therefore have a lesser discrepancy between the above two turbulent disks compared to those around solar-mass stars.

To summarize, the disk turbulent level is important for the water contents of forming planets. In less turbulent disks, the planets mainly migrate inward and end up with relatively low water fractions. The planets formed at the ice line can still be rock-dominated (with a few percent water) when the turbulent $\alphat$ is $10^{-4}$.

\section{ Planet Population Synthesis Modelling and  Observational Comparison}
\label{sec:obs}
We hereafter model the growth and migration of a large number of planets by a Monte Carlo method. The distributions of the disk parameters are used as varying initial conditions described  in \se{setup}. The resulting planet population are then compared with the  observational data. Specifically, the water contents of the planets, the correlation between the planet masses and the masses and metallicities of their stellar hosts are investigated in \se{comparison}.

\subsection{Monte Carlo simulations setup}
\label{sec:setup}

\begin{figure*}[tbh]
 \includegraphics[scale=0.8, angle=0]{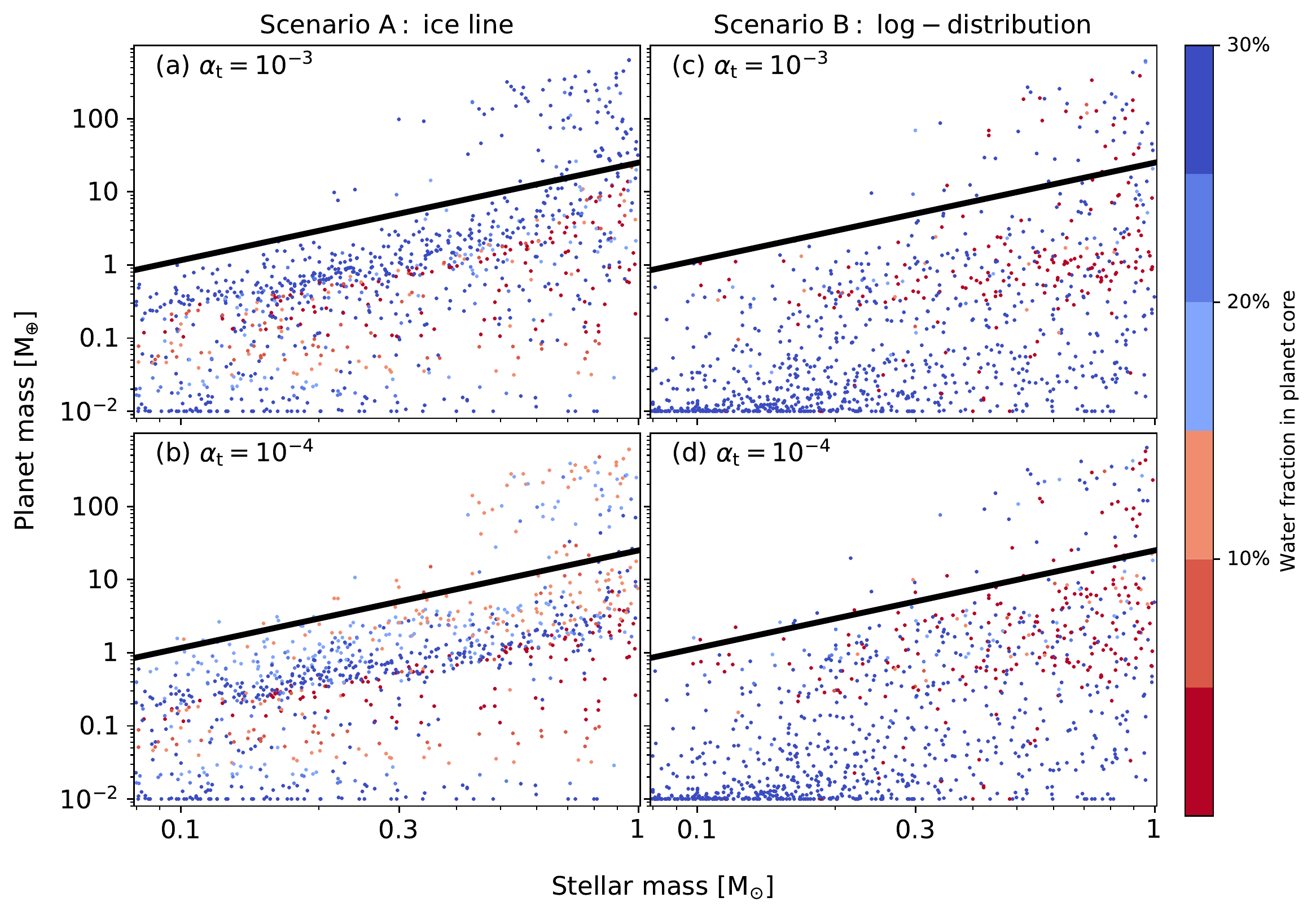}
       \caption{  
              Monte Carlo sampling plot of the planet mass vs the stellar mass, with the ice line planet formation model (Scenario A) in the left, the log-uniform distributed planet formation model (Scenario B) in the right, $\alphat=10^{-3}$ in the top and $\alphat=10^{-4}$ in the bottom.  The color corresponds to the  water mass fraction in the planetary core.  The black line represents the characteristic mass of the simulated super-Earths, set by the pebble isolation mass from \eq{M_iso_vis}. The $M_{\rm p}-M_{\star}$ scaling is almost insensitive to the explored disk turbulent $\alphat$ values and embryos' birth locations $r_0$. The planet with moderate water fraction can form from the water ice line in less turbulent disks. Planets formed over a wide range of disk distances end up with a distinctive, bimodal water mass distribution.
    }
\label{fig:mass}
\end{figure*}

Since the early disk properties are poorly constrained, we use the Monte Carlo approach to generate the varying initial conditions for the planetary formation process. The key parameters ($\dot M_{\rm g0}$, $R_{\rm d0}$, $\xi$, $t_0$ and $M_{\star}$) described as follows are sampled as either gaussian or uniform distributions. The setup of the model parameters are listed  in \tb{distribution}. 

We assume that  the initial gas accretion rate $\dot M_{\rm g0}$ follows a lognormal  distribution (${\rm log} \dot M_{\rm g0} $ follows a Gaussian distribution) with a mean value of $6 \times 10^{-8} (M_{\star}/\Ms)^{1.8} \Msyr$ and a standard deviation $\sigma$ of $0.3$.  
The initial characteristic disk size $R_{\rm d0}$ is assumed to be uniformly distributed from $20$ AU to $200$ AU. 
The pebble to gas flux ratio $\xi$ is adopted uniformly from $0.0033$ to $0.03$ in logarithmic space. We assume the stellar metallicity is a proxy of the solid abundance in protoplanetary disks.   A connection between the relatively pebble abundance $\xi$ (model parameter) and the stellar metallicity $\rm [Fe/ H]$  (observable quantity) is  established by
\begin{equation}
[{\rm Fe/ H}] = \log_{10} (\xi / \xi_{\odot}),
\end{equation}  
where  $\xi_{\odot} =0.01$ corresponds to the disk with solar metallicity $[{\rm Fe/ H}]_{\odot} = 0$. Hence,  $[{\rm Fe/ H}]$ follows a uniform distribution ranging from $-0.48$ to $0.48$. Furthermore,  [{\rm Fe/ H}] also affects the disk opacity since $\kappa$ is contributed from dust grains in protoplanetary disks. We simply assume that $\kappa/\kappa_0=\xi/\xi_{\odot}$. This means that metal rich disks are dusty and more opaque to dissipate energy. 
The embryo injection time $t_0$ is uniformly spanned from the first $0.1$ Myr to $3$ Myr, and the stellar mass $M_{\star}$ is randomly drawn from $0.08 \Ms$ to $1 \Ms$ in logarithmic space.  

Two formation scenarios are considered here. The protoplanetary seeds are either formed only at the water ice line location  (scenario A), or they are uniformly distributed over the entire disk region, from $0.1$ AU to $35$ AU in logarithmic space  (scenario B).  Scenario  A is motivated by the study that at $r_{\rm ice}$ the water vapor could diffuse outwards and re-condense onto the exterior icy pebbles, enriching the local metallicity and triggering the planetesimal formation by the streaming instability \citep{Kretke2007,Ros2013,Ida2016b,Schoonenberg2017,Drazkowska2017}.  In scenario B, planetesimal formation is assumed to be a universal process that can occur at all disk locations.  This could be, for instance that particles are passively concentrated by pressure bumps and collapse into protoplanetary seeds \citep{Lenz2019}.   We assume that this mechanism is not preferred to occur at any particular disk locations. For each scenario, in total $1000$ simulations are performed, with initial conditions randomized from the above distributions.  

Apart from the default setup in \tb{distribution}, we also present the simulations and discuss the influence of these model assumptions and  distributions of parameters in \ap{AP1}, such as the cases with pure stellar irradiation disks, planet growth unlimited by the pebble isolation mass, early formation of the protoplanetary embryos,  a linear stellar mass and luminosity correlation, unchanged disk metallicity during late stellar photoevaporation phase, varied initial disk conditions, and disks with low turbulent viscosities.

\begin{table}
    \centering
    \caption{Adopted parameter distributions for the population synthesis model in \se{obs}}
    \begin{tabular}{lclclclclclc|}
        \hline
        \hline
        Parameter    &   Description  &   \\ 
        disk model &  viscously heated + stellar irradiation \\ 
         $\dot M_{\rm g0}$  ($\Msyr$) & $ 10^{\mathcal {N}(\mu, {\sigma}^{2} ) }$, $\mu=-7.2$ and $\sigma=0.3$ \\ 
         $R_{\rm d0}$ (AU) &  $\rm{U}(20,200)$\\  
         $\xi$ &  $ \rm{\log U }(0.0033,0.03)$ \\ 
          $[{\rm Fe/ H}]$ &  $ \rm{U}(-0.48,0.48)$ \\ 
          $r_0$ (AU) &  Scenario A: the ice line  \\
           &  Scenario B:  $\rm{\log U}(0.1,35)$ \\
           $t_0$ (Myr) & $\rm{U}(0.1,3)$ \\ 
          $M_{\star}$ ($\Ms$)  & $\rm{\log U}(0.08,1)$  \\ 
           $\alphat$   &   $10^{-3}$ and $10^{-4}$ \\ 
           $\alphag$   &   $10^{-2}$ \\ 
           disk opacity   &   $  \kappa = \kappa_0 \left(\xi/\xi_0 \right) $ \\
      \hline
        \hline
    \end{tabular}
    \label{tab:distribution}
\end{table}

\subsection{Models and Observations Comparison}
\label{sec:comparison}

\subsubsection{Planet mass vs stellar mass}
\label{sec:mstar}

The simulated planet populations (scenario A in the left panel and scenario B in the middle panel) are illustrated as a Monte Carlo sampled plot of the planet mass and its stellar mass in \fg{mass}.  The turbulent strength $\alphat$ of $10^{-3}$ and $10^{-4}$ are exhibited in the upper and lower panels.  The color corresponds to the planetary water fraction.

The simulations show that the masses of planets correlate with the masses of their stellar hosts. In both formation scenarios,  planets forming around more massive stars  have higher masses.  Gas giants can  form from these systems whose stellar masses are higher than $0.3\Ms$.  However, for systems around less massive stars ($M_{\star} < 0.3 \Ms$), planets only grow up to a few Earth masses. Furthermore,  when excluding the gas giant population, the upper masses of these small planets also show a clear correlation with their stellar masses, from $1 \Me$ planets orbiting around $0.08 \Ms$ stars to $ 25 \Me$ planets orbiting around solar-mass stars.  There is a narrow concentration of sub-Earth masses  planets in \fg{mass}. This is due to an enhanced disk metallicity and a boosted pebble accretion for low-mass embryos when stellar photoevaporation dominates the disk gas removal.

The above mass correlation can be qualitatively explained as follows.  Planets stop increasing their core masses when they reach the pebble isolation mass.  This isolation mass is lower (higher) for systems around less massive (massive) stars (\eq{m_iso_sim}). For the planets orbiting around stars that are less massive than $0.3 \Ms$, their pebble isolation mass is lower than $5 \Me$. The gas accretion onto such low-mass planets are limited due to a slow Kelvin-Helmholtz contraction.  These planets would eventually strand as rocky or icy planets containing tiny amounts of gas in their atmospheres. The forming planets around more massive stars, however, have a higher pebble isolation mass that is above the critical core mass. These planets, once they reach their pebble isolation mass, can trigger runaway gas accretion and grow into gas giants rapidly.

When excluding the gas giant planets, we find that the upper mass of the small planets (the black line in \fg{mass}) approximately linearly increases with the masses of their stellar hosts. This can be  explained by the stellar mass dependence on their pebble isolation mass. For a simple analysis, we ignore the migration and assume the planet reaches $ M_{\rm iso}$ at $r_{\rm ice}$. From \eqs{h_vis}{rice_vis} the disk aspect ratio at the ice line increases with the stellar mass as $h_{\rm g} \propto \dot M_{\rm g}^{2/9}M_{\star}^{-1/3} \propto  M_{\star}^{1/9}$.  
 Since $h_{\rm g}$ is roughly independent of $r$ in the inner viscously heated region, the $h_{\rm g}-M_{\star}$ scaling does not change too much as the planet migrates.  We can therefore approximate the isolation mass as 
\begin{equation}
 M_{\rm iso} =   25 \left( \frac{ M_{\star}}{1 M_{\odot}} \right) \left( \frac{h_{\rm g}}{0.05} \right)^{3} \Me =  25 \left(   \frac{ M_{\star}}{1 M_{\odot}} \right)^{4/3} \Me.
\label{eq:M_iso_vis}
 \end{equation}  
From \eq{M_iso_vis}, the masses of super-Earths reaching $M_{\rm iso}$ increase slightly super-linearly with their stellar mass.  We use the black line (\Eq{M_iso_vis}) to represent the highest core masses that planets can reach.
 Here the $\alphat$-dependence on  $M_{\rm iso}$ is neglected since $M_{\rm iso}$ only changes by $25\%$  when varying $\alphat$ by one order of magnitude  from $10^{-3}$ to $10^{-4}$ (\eq{m_iso}).  The upper mass of super-Earths  is almost identical in Fig. \fgnum{mass}a and \fgnum{mass}b.

 It is worth noting that  the simulated planets are all in single planetary systems, while  many observed exoplanets are in multi-planetary systems.  After reaching pebble isolation masses, these planets in multiples could become dynamically unstable and grow their masses through giant impacts. This process, which is most likely occurred after the gas disk dispersal,  is not taken into account in this work (\eg , see \citealt{Ogihara2009,Izidoro2017}). Thus, for planets in multiples, our pebble isolation mass can be treated as a lower limit for the final planet mass \citep{Izidoro2019,Lambrechts2019}.

 Now let us look back to observations (\fg{massobs}). Although the number of gas giants seems to be more than that of the super-Earths in \fg{massobs}, this is due to the detection bias, where low-mass planets is in general difficult to detect than massive planets. Statistically, the occurrence rate of close-in super-Earths is $\approx$$30\%$ \citep{Fressin2013,Zhu2018} while $\approx$$10-15 \%$ of solar-type stars have gas giants  with orbits out to a few years \citep{Cumming2008,Mayor2011,Fernandes2019}. The simulated planet population (\fg{mass}) roughly matches this ratio, for both Scenario A and B.  
 As stated in \se{introduction},  the observed $M_{\rm p}-M_{\star}$ trend is not owing to the observational bias. 
In addition, a similar linear scaling of the characteristic mass on the stellar mass for the observed Kepler planets is also obtained by \cite{Wu2018} and \cite{Pascucci2018}, using different probing methodology.  Our simulated planet population thus agrees well with the trend shown in exoplanet observations.

As a key remark, we propose  that the characteristic planet mass is related to the fact that the growing planet generates a  gas pressure bump at its surrounding, which halts the further core growth by pebble accretion.  
Therefore, the representative of the characteristic mass is the pebble isolation mass, which linearly scales with the stellar mass.

\subsubsection{Water content}

Disk turbulence affects both the efficiency of pebble accretion and migration direction. We find in \fg{mass} that planets formed  in lower turbulent disks generally contain a lower water fraction. This is because the outward migration is limited in this case and the planets has spent less time outside of $r_{\rm ice}$ accreting wet pebbles. For the water ice line formation scenario, the water fraction of the most massive super-Earth planets is close to $30 \%$  in the disks of $\alphat =10^{-3}$ (\fg{mass}a), while  the water fraction reduces to $10\%-15\%$ level in the disks of  $\alphat =10^{-4}$ (\fg{mass}b). As will explain later, the latter range of water fractions in fact agrees better with the inferred composition of the Kepler planets.  For scenario B,  the assumed random starting points allow the resulting super-Earths to be very water-rich ($f_{\rm H2 O} >30\%$) or water-deficit ($f_{\rm H2 O} <5\%$), depending their birth locations relative to the ice line.  The finding water fractions in the cores of gas giant planets are similar compared to  super-Earths in both scenarios.

We note that in Scenario A  embryos are assumed to contain $35\%$ water at the ice line initially.  On the other hand, \cite{Ida2016} and \cite{Hyodo2019} suggested that dry planetesimals could also be generated slightly interior to the ice line, due to a `traffic jam' effect between the slow drifting silicate grains and fast drifting icy pebbles. The circumstance of embryos formed slightly interior to  the ice line with initial zero water fraction is also tested. We find that the result is quite similar to \fg{mass}a, especially for the planets whose masses are higher than $0.1 \Me$. Since the inward movement of the ice line is faster than the migration of low-mass embryos at the beginning, in both cases the growing embryos would soon accrete wet pebbles. These embryos would eventually migrate interior to the ice line and accrete dry pebbles when they become substantially massive.  As a result, in the ice line formation scenario the total water content of final planets is insensitive to their initial values, but really depends on how long they can retain outside of $r_{\rm ice}$.

 \cite{Ormel2017}  proposed a scenario of forming planets at the water ice line can explain the formation and compositions of TRAPPIST-$1$ planets. \cite{Schoonenberg2019} later carried out numerical simulations and found the water content of resulting planets is $\sim$$10\%$ based on the above scenario. There are a few differences between their work and ours. First,  in  \cite{Schoonenberg2019}'s model pebble size is determined by the radial drift and fragmentation while in our model pebbles are smaller, assuming that they are limited by bouncing. This results in a high pebble flux  and a rapid planet formation during early stage in their case.
 Second, only inward migration is considered in \cite{Schoonenberg2019}, while this work also accounts for the effect of outward migration, which depends on  the turbulent viscosity. 

Recently,  the California-Kepler Survrey reported  a factor of two drop in the occurrence rate of planets with radii  $\sim$$1.5 R_{\oplus} - 2 R_{\oplus}$ \citep{Fulton2017,Fulton2018}.  This planetary radius valley indicates a composition transition from the terrestrial-like, atmosphere free planets to the planets with gaseous envelopes of a few percent in mass.  Such super-Earth planets  formed early in disks would have accreted  primordial $\rm H/He$ envelopes.  The envelope mass is lost afterwards either due to the external stellar  photoevaporation \citep{Owen2017,Jin2018} or the internal core-powered heating \citep{,Ginzburg2018,Gupta2019}.   

The composition (bulk density) of the observed planet population can in principle be constrained, but requires substantial knowledge of the exact location of this valley, the relative variation of the occurrence rate within/out of the valley,  and detailed modelling the planetary internal structure.  For a first attempt, \cite{Owen2017} proposed that the current data rules out  very low density, water world planets of $\rho_{\rm p} \simeq 1.5 \ \rm g \ cm^{-3}$, and   the terrestrial-like composition of $\rho_{\rm p} \simeq 4 \ \rm g \ cm^{-3}$ are more preferred.
\cite{Gupta2019} obtained a similar result and further constrained that the water fraction for Earth-like composition cores can be up to $ 20\%$, although in their model the envelope mass is lost due to luminous cores rather than stellar photoevaporation. 
The radioactive heating is not taken into account in these previous studies. Notably,  \cite{Vazan2018} found that the long-term thermal effects (\eg , radioactive decay, latent heat from the core solidification) on the core-envelope evolution can contribute to the variation of the radii up to $ 15\%$.  There might still be a considerable degeneracy on the internal core heating sources and the chemical compositions. Probing the bulk density of the planetary cores  from this radius valley therefore needs to be treated cautiously.  To be conservative, the inferred close-in super-Earths could still contain core water mass fraction up to $10\%-20\%$.  Altogether, the planets generated by our model in lower turbulent disks are consistent with the observed Kepler planets. 

After the formation phase and gas disk dissipation, the planets experience the runaway greenhouse effect where they can lose substantial amount of the surface water. First, strong stellar FUV  radiation photo-dissociates $ \rm H_2O$ molecule. The  X-ray and ultraviolet (XUV) irradiation drive the hydrodynamic escape of hydrogen, and potentially some of oxygen as well \citep{Luger2015}.  Meanwhile, the water formed in the mantle of the planet could subsequently be released into the atmosphere  through evaporation of the water mantle or outgassing from volcanoes.
It is important to note that the above water loss process is significant for systems around low-mass stars compared to more massive stars, since these low-mass stars are more magnetic active due to a slow contraction and cooling.  
\cite{Bolmont2017} applied this scenario for TRAPPIST-$1$ system and found that the two planets in habitable zone, TRAPPIST-$1$b and c may lost  $\approx 10$ Earth oceans. 
\cite{Tian2015} included this late phase water loss in their planet population synthesis model. Their result shows that Earth-like planets in the habitable zones around M-dwarfs ($M_{\star} = 0.3 \Ms$) eventually turn into either water-rich, ocean planets or entirely desert planets, depending on their initial water content.

\subsubsection{Planet mass vs stellar metallicity}
\label{sec:Zstar}

\begin{figure*}[t]
  \includegraphics[scale=0.8, angle=0]{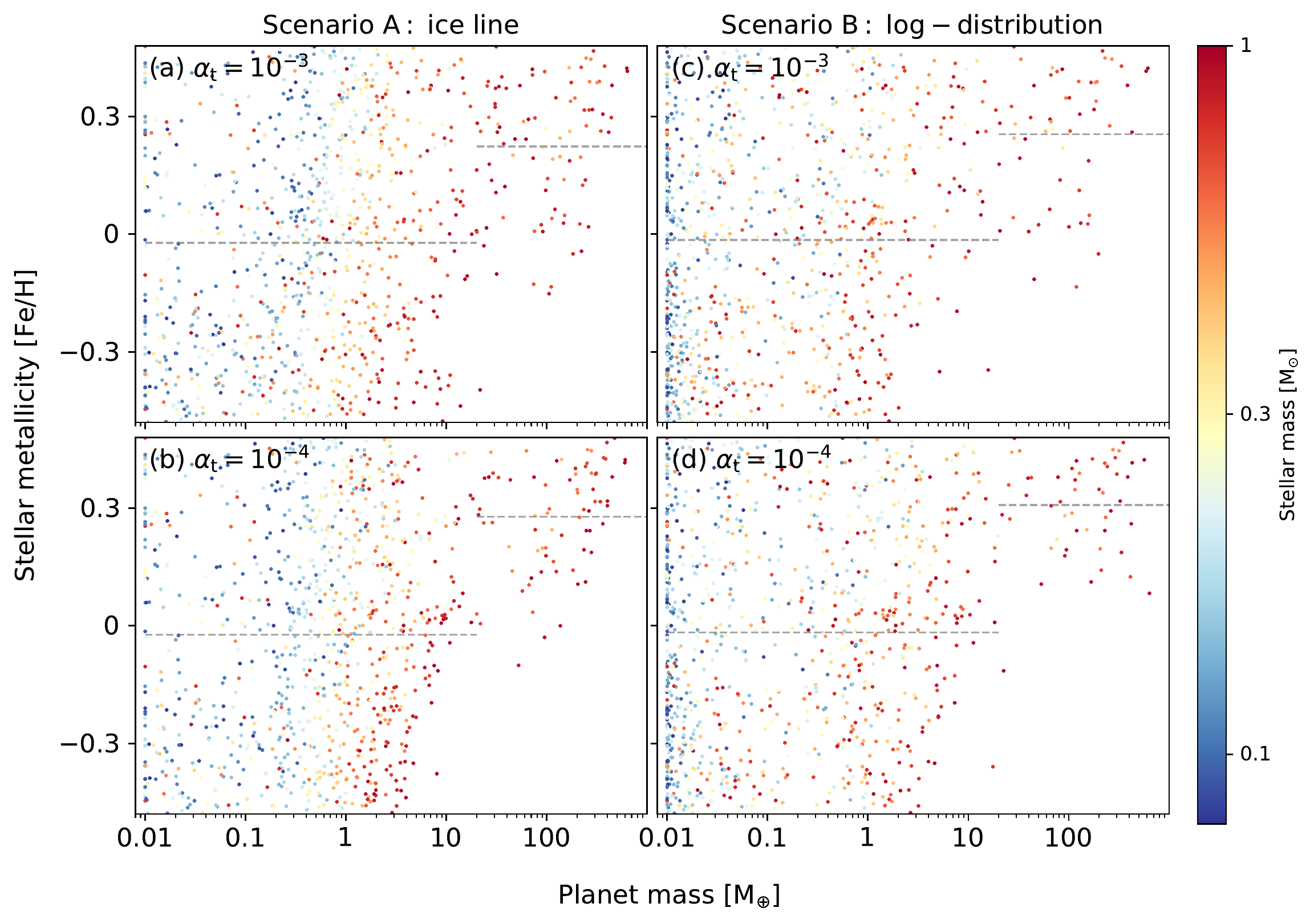}
       \caption{ 
          Monte Carlo sampling plot of the planet mass vs the stellar metallicity, with the ice line planet formation model (Scenario A) in the left, the log-uniform distributed planet formation model (Scenario B) in the right, $\alphat=10^{-3}$ in the top and $\alphat=10^{-4}$ in the bottom.  The color corresponds to the stellar mass.  The grey dashed lines represent the averaged metallicities for the planets with $M_{\rm p} \leq 20 M_{\oplus}$ and $M_{\rm p} >20 \Me$. The formation of low-mass planets ($M_{\rm p} \leq 20 M_{\oplus}$) is barely dependent on the stellar metallicity,  whereas massive planets ($M_{\rm p} >20 \Me$) prefers to grow in metal-rich systems.  
       }
\label{fig:metallicity}
\end{figure*}

\begin{figure*}[t]
 \includegraphics[scale=0.8, angle=0]{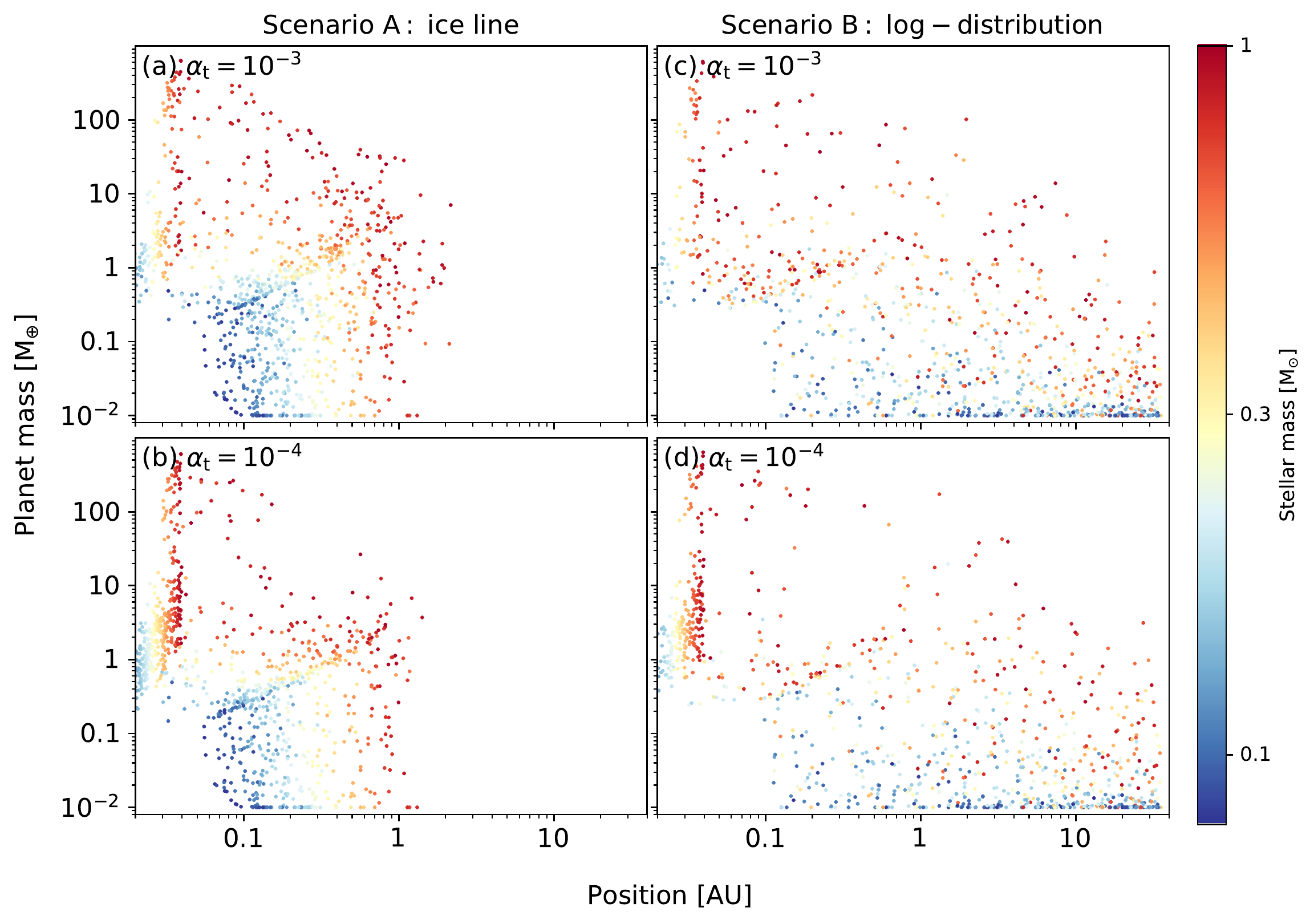}
       \caption{ 
  Monte Carlo sampling plot of the planet mass vs the semimajor axis, with the ice line planet formation model (Scenario A) in the left, the log-uniform distributed planet formation model (Scenario B) in the right, $\alphat=10^{-3}$ in the top and $\alphat=10^{-4}$ in the bottom.  The color corresponds to the stellar mass.  Scenario A only produces close-in planets,  while planets formed in Scenario B have a wide range of orbital distances.}
\label{fig:rplanet}
\end{figure*}

\fg{metallicity} shows the the planet mass as a function of the stellar metallicity. 
 The color corresponds to the mass of the central star. Both two scenarios and two different $\alphat$ show a similar trend that metal rich stars have more large planets, particularly for gas giant planets.  For instance,  planets with $M_{\rm p} \gtrsim50 \Me$ are more frequently occurred in metal rich systems. These planets are rare in systems of $[{\rm Fe/H}] \lesssim -0.3$. On the other hand, the metallicity dependence on the low-mass planets is quite insignificant.  One order of magnitude change in the disk pebbles mass (proportional to $\xi$) leads to approximately a factor of two change in upper mass of super Earth planets in \fg{metallicity}.  
 
 The upper mass of these super Earths is again determined by the pebble isolation mass $M_{\rm iso}$. Although $M_{\rm iso}$ does not not directly correlate with $\xi$,  the planets reach the isolation mass at early time in the metal rich systems where disks contain higher pebble fluxes.    On average, the growing super-Earths around metal rich stars have more time to accrete sufficient gas to become gas giant planets while they remain gas poor around metal-deficient stars.  
  In the inner viscously heated region $h_{\rm gas}$ only weakly depends on  the disk opacity (\eq{h_vis}) and therefore the metallicity. As a result, the (upper) mass of super-Earth planets shown in \fg{metallicity} modestly dependent on the metallicity of their stellar hosts.

 Observations indicate that the frequency of the gas giants increases with stellar metallicity \citep{Santos2004,Fischer2005,Sousa2011} while the super-Earths are commonly detected around stellar hosts with a wide range of metallicities \citep{Mayor2011,Wang2013,Buchhave2014}. \cite{Buchhave2014} divided planets detected by the Kepler satellite  into different radius bins and found that the mean metallicity of their stellar hosts correlate with the planet radius.  In their Fig.1 the mean metallicity  of super-Earths with $R_{\rm p} \leqslant 4R_{\oplus}$ is around solar value, while that of gas giant planets with $R_{\rm p} > 4R_{\oplus}$ is close to $0.2$. Here we also divide our planets into two samples, low-mass planets with $M_{\rm p} \leqslant 20 \Me$ and massive planets with  $M_{\rm p} >20 \Me$. The mean metallicities of these two samples are represented by grey dashed lines in  in \fg{metallicity}. We find that the metallicity correlation from the simulated planets is consistent with the observations.     
In short, the strong metallicity dependence on the formation of gas giant planets but not super-Earths appears to be the key feature for pebble accretion scenario.   

\subsubsection{Planet mass vs semimajor axis}
\label{sec:rplanet}

\fg{rplanet} shows the Monte Carlo plot of the planet mass and its semimajor axis. The color represents the mass of its stellar host. We can see the growth and migration of  the planets around stars of different masses. Scenario A, by default starting the growth of the embryos at the water ice line, only produces close-in planets with orbits $\lesssim 1 \AU$. Distant planets can form when their birth locations are further out.   Planets generated from Scenario B are eventually distributed over a wide range of disk regions.

Observations imply that  the occurrence rate of cold gas giants is a factor of $5-10$ higher than the hot-Jupiters \citep{Cumming2008,Wright2012}.   One caveat is that even scenario B produces too many hot Jupiters and too few cold gas giant planets.  In other word, the migration is still too fast. 
Lower turbulent disks (smaller $\alphat$) on the one hand could help because in this case pebble accretion is faster.  Planets spend less time to reach the same masses, and therefore the radial migration distance is shorter. However, on the other hand, the outward migration effect is weaker in lower turbulent disks.  The $\alphat =10^{-4}$ case shown in \fg{rplanet}d is still not enough to fit the observations.

For the cold gas giant planets ending up with a few AU orbits, their initial seeds  should form very far away.
As already  indicated by \cite{Johansen2018}, if the sublimation fronts of other volatile species (\eg , the $\rm CO$ ice line at $30$$-$$40 $ AU) are equally prone to form planetesimals as the water ice line, these seeds could potentially grow into cold gas giants. 

Here we only consider the growth of one single embryo for a given system. The evolutional outcome can be different if multiple embryos form together and the mutual dynamical interactions are taken into account.  These embryos sequentially migrate toward to the transition radius and get trapped in compact resonances.  Such a mass concentration may trigger dynamical instability even when the gas disk is still presented. Some planets are occasionally scattered to larger orbital distances and grow massive there.  These planets which originally form hot Jupiters in a single planet formation picture, may instead grow into cold gas giant planets in this case.  The above mentioned process will be studied with a N-body code in a future work.   The other speculation is that a large fraction of the forming hot Jupiter cannot survive long enough. They could be swallowed by the central stars via strong tidal interactions and subsequent Roche lobe overflow \citep{Trilling1998,Gu2003,LiuSF2013}. In addition, the stellar photoevaporation strips the envelopes of very close-in gas giants, turning (a fraction of) them into low-mass planets \citep{Lammer2003,Lecavelier2004}.

\section{Conclusions}
\label{sec:conclusion}

In this paper we have constructed a semi-analytical  population synthesis model based on  the pebble-driven planet formation scenario. A two-component (viscously heated + stellar irradiation) gas disk structure is taken into account. The evolution of the disk is driven by viscous accretion for most of the time, and the stellar X-ray photoevaporation dominates the gas removal in the end. The physical processes of planet formation includes pebble accretion onto planetary cores, gas accretion onto their envelopes, and planet migration (\se{method}). The influence of key disk parameters on the planet growth are illustrated in \se{result}. 
We investigated the final masses and the water contents of the forming planets around solar-mass stars and very low-mass M dwarfs of $0.1 \Ms$ at two disk turbulent strengths  $\alphat$ of $10^{-3}$ and $10^{-4}$ in \se{map}.  Two hypotheses have been explored, assuming that the protoplanetary seeds are either originated only at the water ice line or log-uniformly distributed over the entire disk regions. The planets generated from the population synthesis model are compared with the observations in \se{obs}.  
 
The major findings of this paper are summarized as follows: 
\begin{enumerate}[1.]
     \item The outcome of planet formation depends strongly on the initial disk and stellar properties. Massive planets can grow in circumstances  when the initial characteristic disk size ($R_{\rm d0}$) is larger,  the initial disk accretion rate ($\dot M_{\rm g0}$) is higher, the central star is more massive, and the disk metallicity is higher (\fg{growth}).
     \item  Jupiter-mass  planets are preferred to form in the early phase at the disk region close to the ice line at $\alphat =10^{-3}$ or further out at tens of AUs at $\alphat =10^{-4}$. Earth mass planets can form near the ice line region around  stars of $0.1 \Ms$ (\fg{mapmass}).  
      \item  The  water fractions in planetary cores  reach $35\%$ when they grow far outside of $r_{\rm ice}$. 
      Rocky planets with much less than $1\%$ water fractions  can only form when the embryos grow interior to $r_{\rm ice}$. The water contents of the planets formed at the ice line ranges from $\gtrsim 10\%$  to  $1\%$, depending on the disk turbulence and the masses of the stellar hosts. Water-deficit planets are more likely to form  when the disks are  less turbulent  and/or their stellar hosts are more massive (\fg{mapwater}).       
     \item  The characteristic planet mass is set by the pebble isolation mass, which increases approximately linearly with the stellar mass  (\eq{M_iso_vis}).  
     \item For the ice line formation model, the  super-Earth planets can have $\sim$ $10-20\%$ of water mass in their cores at $\alphat = 10^{-4}$, while they end up with a higher water content of $\gtrsim 20\%$ at  $\alphat = 10^{-3}$. 
     For the log-uniformly distributed formation model, planets end up into a bimodal composition distribution, being either water-rich ($f_{\rm H_2 O} \simeq 35 \%$) or substantially dry  ($f_{\rm H_2 O} < 1 \%$).   
       \item Core-dominated planets with a mass lower than the characteristic mass can form in systems with a wide range of stellar masses and metallicities. Nevertheless,  gas-dominated planets with a mass higher than the characteristic mass can mainly form when the central stars are more massive than $0.3\Ms$ (\fg{mass}), and/or the stellar metallicities are higher than $-0.15$ (\fg{metallicity}). These simulated features are in good agreement with the observed exoplanet population.     
 \end{enumerate}
 
Overall, the key conclusion in our study is that the characteristic core-dominated planet mass may be set by the pebble isolation mass, when the feeding of pebbles is terminated by a gap-opening planet.  Our proposed linear correlation between the characteristic planet mass and the stellar mass is consistent with the findings by \cite{Wu2018} and \cite{Pascucci2018}.
  We here use planetary mass rather than radius. This is because, in contrast to the planet mass,  the radius may change significantly due to the envelope loss by stellar photoevaporation \citep{Owen2017,Jin2018} or the contraction of luminous cores \citep{Gupta2019}.  One drawback to use the mass  is that the masses of most transiting planets (such as the planets detected by the Kepler satellite) are not well constrained.  This limits the observational sample for a statistical comparison.  The ongoing or upcoming missions, like Mearth project, SPECULOOS (Search for Habitable Planets Eclipsing Ultra-cool Stars),  TESS (Transiting Exoplanet Survey Satellite) and  PLATO  (Planetary Transits and Oscillations of stars) will  provide great opportunities to detect planets around nearby bright stars. The masses of these planets can be following measured  by  ground-base radial velocity surveys accurately.
The target stars of previous ground based and Kepler surveys are mostly FGK and early M dwarf stars.  These missions will also be able to detect a large number of planets around very low-mass (late) M-dwarfs.  Consequently, a more precise and large data sample over a wide range of stellar masses will help to further verify the correlations predicted by our model.

\appendix

\section{Influence of model parameters}
\label{ap:AP1}
We test a few parameters and model assumptions here and discuss how the resulting planet populations differ from the simulations shown in \se{obs}.

\subsection{ Stellar irradiated disk}
\label{ap:irradiation}
In order to test the importance of viscously heated disk regions for the formation of planets, we perform additional simulations considering only stellar irradiated disks with seeds that are started from the ice line (Scenario A). The other parameters remain the same as in \tb{distribution}. From \eq{T_irr} to \eq{h_irr}, the corresponding temperature and scale height in such a disk remain constant during the disk evolution.  

\fg{appendix1} illustrates the Monte Carlo plot of the planet mass as a function of the stellar mass for both $\alphat =10^{-3}$ and $\alphat= 10^{-4}$.  Although  there is still a mass correlation between planets and  their stellar hosts, the resulting planets in \fg{appendix1} have much lower masses compared to those formed in disks when the inner viscous heated regions is included (\fg{mass}). There is no any gas giant planets in \fg{appendix1}. 
 The most massive planet reaches only $2\Me$ around  solar-mass stars, while it attains $0.1 \Me$ around  $0.2 \Ms$ stars.

From \eqs{h_irr}{rice_irr}, the disk aspect ratio at the ice line  in a stellar irradiation disk $h_{\rm g, irr} \propto M_{\star}^{-2/3}L_{\star}^{1/3}$. Adopting  $L_{\star} \propto M_{\star}^2$  we find that $h_{\rm g, irr}$ is independent of the stellar mass. Therefore, the isolation mass at the ice line linearly increases with the stellar mass, $M_{\rm iso} \simeq 2 \Me (M_{\star}/M_{\odot})$. There are a few differences when the viscous heating is taken into account. First, the planet can migrate outward and temporally stall at $r_{\rm tran}$.   The time span for accreting materials is longer in this case. Second, $h_{\rm g, vis}$ is larger in disks of  a higher accretion rates, and therefore the isolation mass in the viscously heated region can be higher than that in the stellar irradiation region. Third, the feature of viscously heated disk region is that $h_{\rm g, vis}$, and hence $M_{\rm iso, vis}$, are almost independent of the position of the planet.  
On the contrary, in pure stellar irradiated disks $h_{\rm g, irr}$ and $M_{\rm iso,irr}$ are lower when $r$ is smaller. Thus, the inwardly migrating planet with a shorter orbit ends up into a lower pebble isolation mass.  
Altogether, these factors eventually lead to  the formation of more massive planets in the two-component  disks (\fg{mass}) while only low-mass planets appear in pure stellar irradiation disks (\fg{appendix1}).  The upper limit of  the planet masses shown in  Fig.\fgnum{appendix1}a and \fgnum{appendix1}b are not very different. This is because $M_{\rm iso}$ does not vary too much when $\alphat $ changes from $10^{-3}$ to $10^{-4}$.    In order to form more massive planets (\eg , approaching the black line in \fg{appendix1}) in such stellar irradiated disks, extensive giant impacts among these protoplanets during or after the gas disk dispersal are essential.

We also find that more massive planets have a lower water fraction   in \fg{appendix1}.  This is because  these massive planets have migrated over larger radial distances inside $r_{\rm ice}$, and therefore accrete more dry pebbles. The key reason for the formation of water-rich planets  in \fg{mass} is that these planets can migrate outward in the viscously heated region and grow a  significant fraction their masses by accreting wet pebbles beyond $r_{\rm ice}$.

\begin{figure}[t]
 \includegraphics[scale=0.7, angle=0]{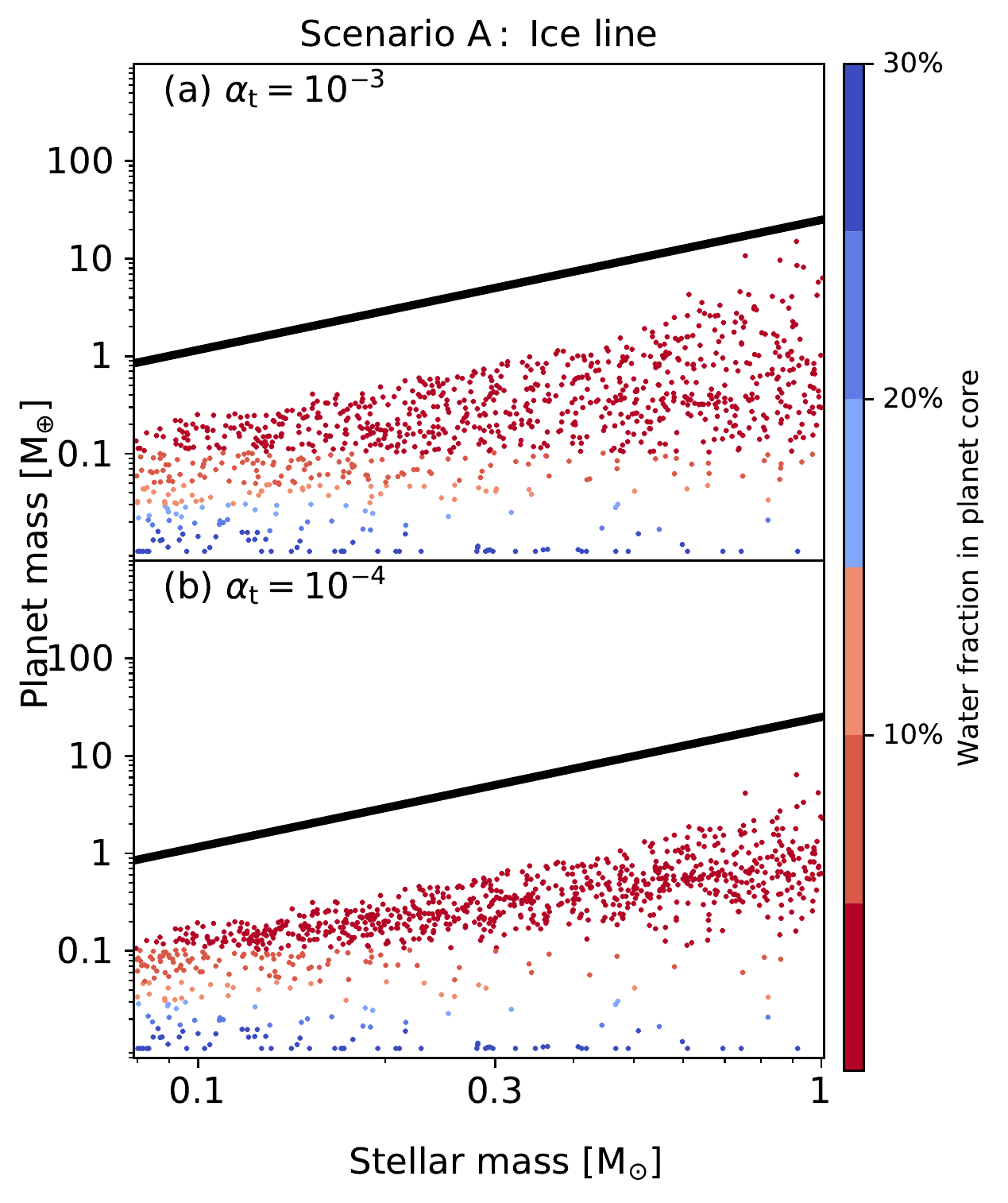}
       \caption{ 
       Same as Fig. \fgnum{mass}a and \fgnum{mass}b, but the disks are only stellar irradiated. The masses of resulting planets are generally lower than \fg{mass}, but also linearly scale with their stellar masses. The massive planets are rocky dominated while  the low-mass planets contain $\gtrsim 20\%$ water. 
    }
\label{fig:appendix1}
\end{figure} 

\begin{figure}[t]
 \includegraphics[scale=0.7, angle=0]{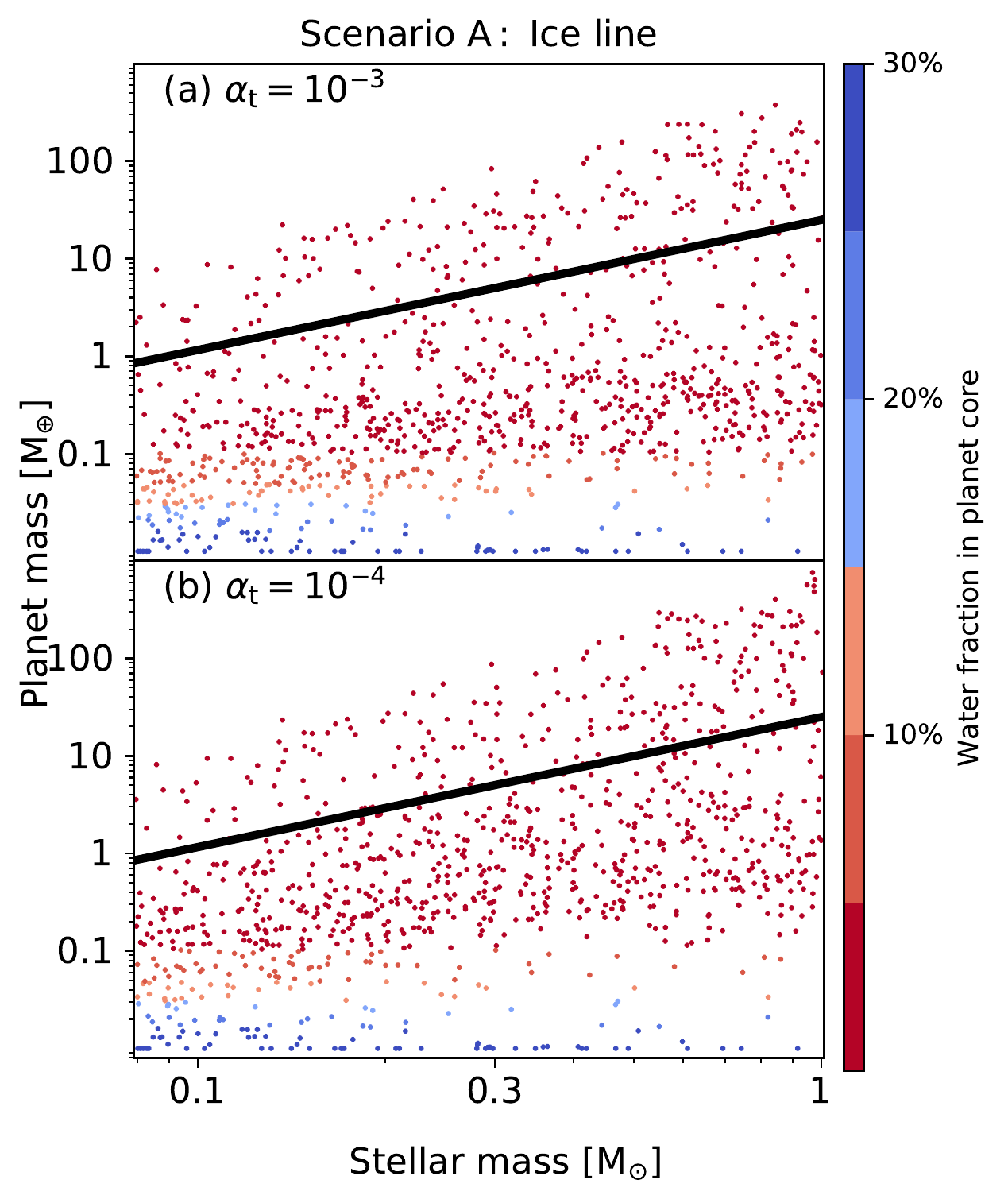}
       \caption{ 
  Same as \fg{appendix1}, but the planets are allowed to accrete pebbles when their masses are higher  than the the pebble isolation mass.  Planets growth truncated by  migration also shows a similar linear scaling.     
    }
\label{fig:appendix2}
\end{figure}

\subsection{Planet mass truncated by migration}
\label{ap:migration}
We consider one idealized situation when the planet core mass is not limited to the pebble isolation mass. In this case planets only terminate accretion when they enter into the inner disk cavity. For a direct comparison with \ap{irradiation}, we also adopt a pure stellar irradiation disk and therefore planets only migrate inward.    We define the final planet mass that is limited by migration as $M_{\rm mig}$, which differs from  $M_{\rm iso}$ when inwardly drifting pebbles are truncated by a planet-induced gap. 

Comparing to \fg{appendix1}, we find that more massive planets form in this migration limit regime. For instance, a few Earth mass planets can form around very low-mass stars in \fg{appendix2} while only $0.2 \Me$ planets appear in \fg{appendix1}.  But  it is still unlikely to grow massive cores and form gas giant planets around stars of $\lesssim 0.1 \Ms$.  In addition, more planets with a very low water fraction (dark blue) are shown in \fg{appendix2}.    

 From the adopted disk and stellar parameters, we find that $M_{\rm mig} >M_{\rm iso}$.  Planets that reach  $M_{\rm iso}$ can thus continue to accrete dry pebbles until they enter into the inner cavity. Therefore, planets would end up with a low water fraction.   We derive that $M_{\rm mig}/M_{\star} \propto h_{\rm g}^{1.5}$ when only accounting for the type I migration and pebble accretion.  As a result, we can see that  $M_{\rm p}$ also increases linearly with  $M_{\star}$ in \fg{appendix2}.

Again we note that this setup is not realistic,  since planets would stop pebble accretion when they reach $M_{\rm iso}$ where the gravity of the planet is strong enough to reverse the local disk pressure gradient, halting the inward drifting pebbles.  The lesson we learn is that, even though there would be no $M_{\rm iso}$  constraint, the masses of the growing planets would still be limited by migration.  We also would like to point out that the planet migration behavior and history can be far more complicated. For instance, additional torques could be generated when considering planet accretion \citep{Benitez-Llambay2015} and dust dynamics \citep{Benitez-Llambay2018}.  The resulting  $M_{\rm mig}$, and therefore the mass scaling may differ from our simple treatment shown in \fg{appendix3}. Nevertheless, the planet mass determined by $M_{\rm iso}$ exhibits a relative clean stellar mass dependence.

We assume that the planets directly migrate into the inner cavity and stop pebble accretion. Another plausible situation is that the low-mass planets may stall at the outer edge of the magnetospheric cavity \citep{Liu2017,Romanova2019} and keep accreting pebbles. In this case the pebble accretion will be terminated when planets reach $M_{\rm iso}$ but not $M_{\rm mig}$. However,  since  $M_{\rm mig} >M_{\rm iso}$ for the adopted disk and stellar parameters in our work,  the planets already attain the pebble isolation mass before reaching the inner cavity. Even they can stall at the disk edge, no further pebble accretion would proceed.  Therefore, whether planets directly enter the cavity or stall at the edge of the cavity would not make a difference in our model.

\subsection{Early seeds formation}
\label{ap:early}

  In \se{obs} we assume a distribution of ejection time of the protoplanetary embryos ($t_0$) from $0.1$ Myr to $3$ Myr.  The other hypothesis is that the first generation of planetesimals formed very rapidly, $t_0\lesssim 10^{5}$ yr. This might be supported, for instance,  by the age measurements of iron meteorites in our solar system \citep{Kleine2005}. In order to test this early formation scenario,  we artificially set $t_0 = 0$ yr and the other parameters remain the same as in \tb{distribution}. Here we consider scenario B and the resulting  planets are illustrated in \fg{appendix3}.   

 Compared to Fig. \fgnum{mass}c and \fgnum{mass}d,  we find that the mass growth is more significant  in \fg{appendix3}.  Here a higher fraction of embryos grow into super Earths and gas giant planets.  The upper mass of super Earths also increases, due to the fact that embryos attain $M_{\rm iso}$ at a higher $h_{\rm gas}$ in early stage.  As a result, the formation time of the embryos affects the transition mass from  low-mass planets to massive giant planets, and essentially the overall mass distribution of the planet population. We also find  more rocky planets in \fg{appendix3} than those in Fig. \fgnum{mass}c and \fgnum{mass}d. This is because the ice line is further out at early stage (\fg{mapmass}), and a higher fraction of embryos in Scenario B are initially inside of the water ice line in this case.

\begin{figure}[t]
 \includegraphics[scale=0.7, angle=0]{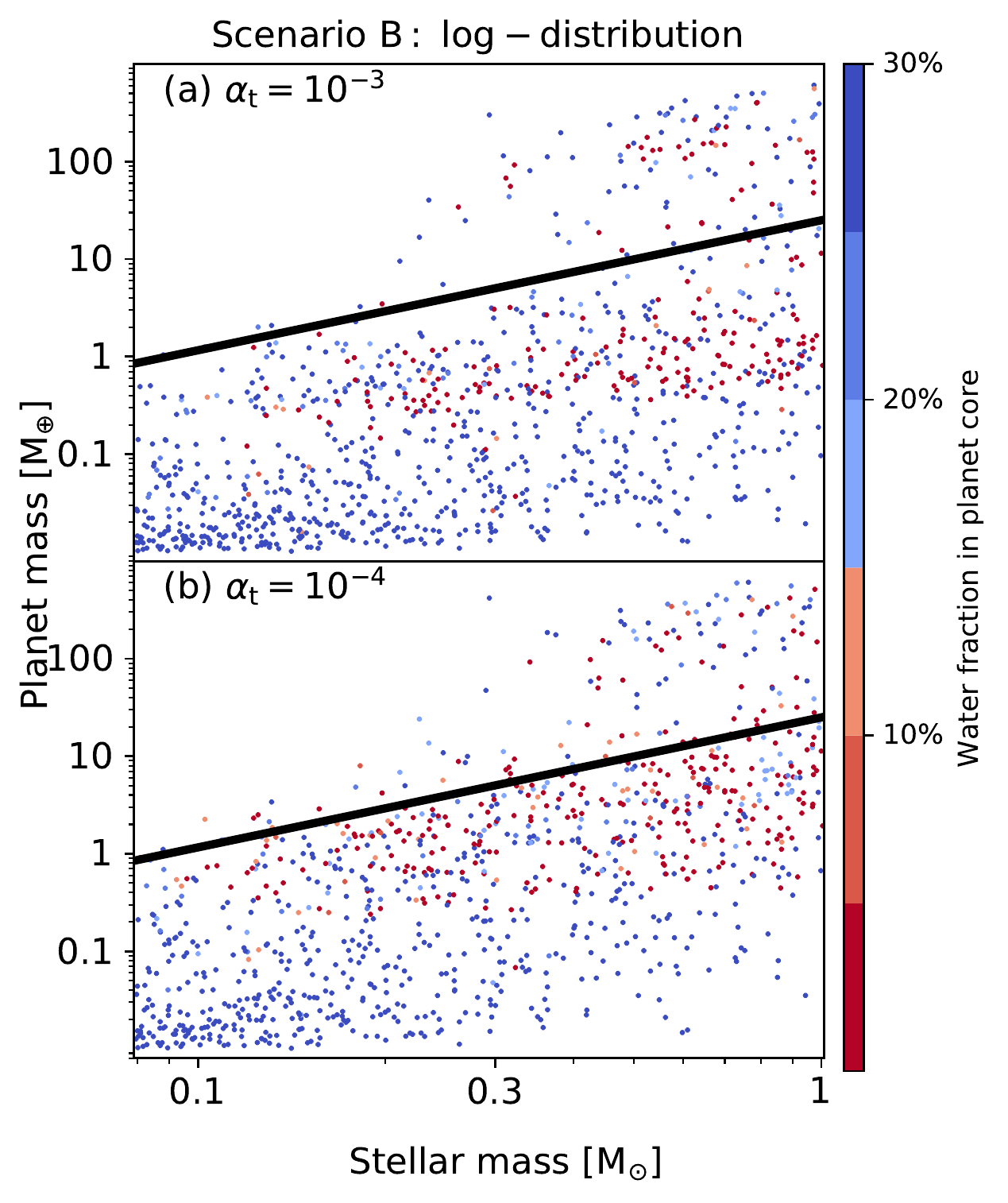}
       \caption{ 
        Same as Fig. \fgnum{mass}c and \fgnum{mass}d,  but  all embryos are assumed to form early at $t_0 =0$ yr. The early formation embryos have a more substantial mass growth. The planet mass distribution is crucially related with the seeds formation time.} 
\label{fig:appendix3}
\end{figure}

\begin{figure*}[t]
 \includegraphics[scale=0.7, angle=0]{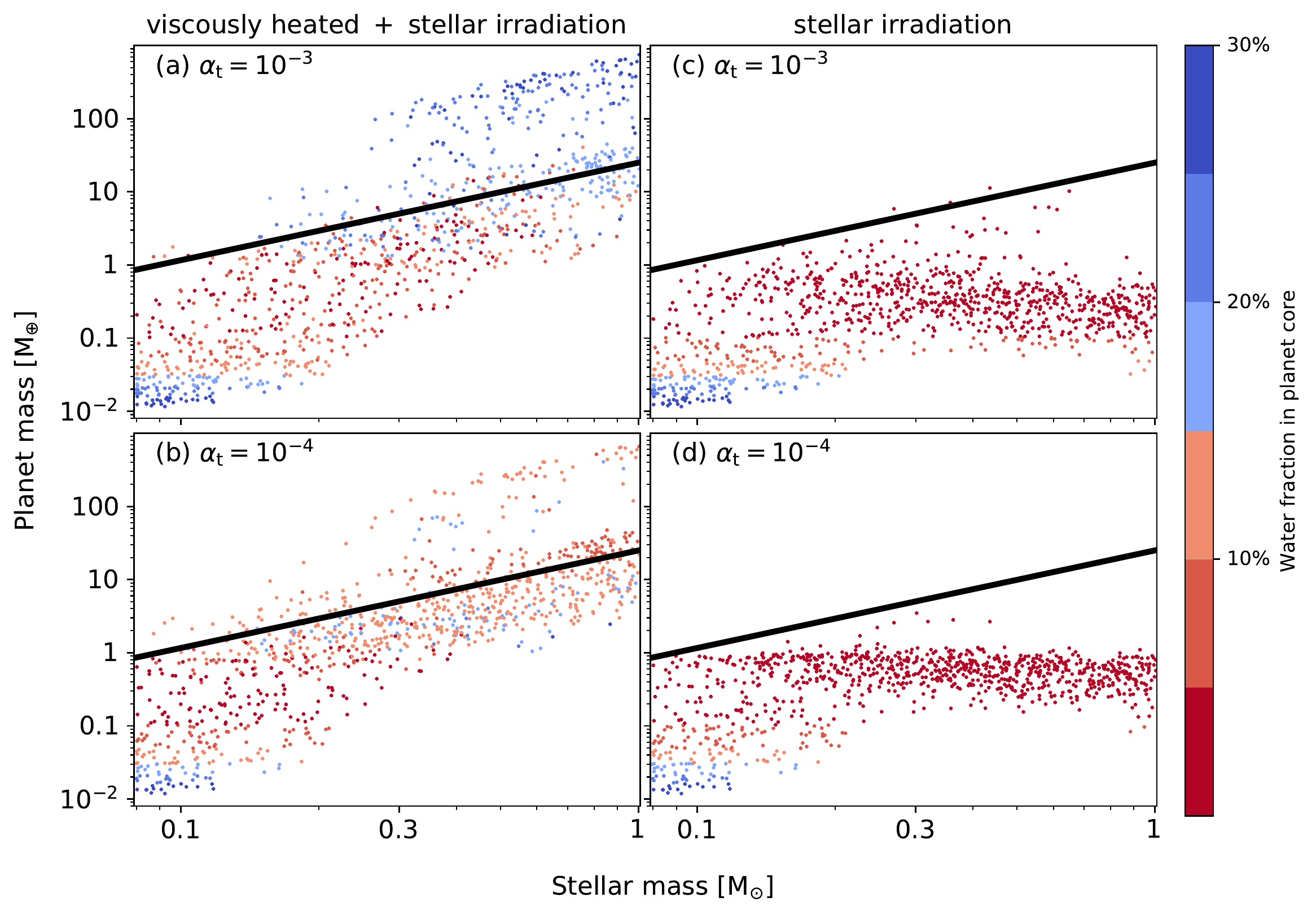}
       \caption{ 
        Same as Fig. \fgnum{mass}a and \fgnum{mass}b,  but  a linear stellar mass and luminosity correlation is adopted, with viscously heated and stellar irradiation disks in the left panel and pure stellar irradiation disks in the right panel.  The characteristic disk size is fixed at $300$ AU, and all embryos are assumed to start their growth at $t_0 = 0$ yr.   The linear mass scaling between the planet and star can be obtained when the viscously heated inner disk regions are included. The planet masses are independent of their stellar masses when only stellar irradiated disks are considered.   
    }
\label{fig:appendix4}
\end{figure*} 

\subsection{A linear stellar mass and luminosity correlation}
\label{ap:luminosity}

In the main paper we assume that the stellar mass and luminosity follows  $L_{\star} \propto M_{\star}^2$.  As discussed in \se{result}, the luminosity of the young, active stars are largely uncertain. Here we also run simulations when the stellar luminosity linearly correlates with the stellar mass, $L_{\star} \propto M_{\star}$, where the luminosity of a solar-mass star is still adopted to be $ 1 \  L_{\odot}$. In this case solar-mass stars are the same but  low-mass stars are brighter than the previous $L_{\star} \propto M_{\star}^{2}$ correlation. 

 When a shallower $L_{\star}$-$M_{\star}$ correlation is adopted, the low-mass stars become brighter and  disks are hotter via stellar irradiation. Hotter disks have larger viscosities and therefore evolve more rapidly. These disks  have a shorter lifetime and the heating is quickly dominated by stellar irradiation compared to disks in \fg{mass}.    In oder to maintain the inner viscously heated region for planet growth, we adopt large disks of $R_{\rm d0} = 300$ AU and allow the embryos to grow early at $t_0= 0$ yr.  Planets therefore have sufficient time to grow their masses before disk gas is all depleted.  Apart from these two-component disks,  we also consider the disks with only stellar irradiation. The other model parameters are the same as in  \tb{distribution}. 

 \Fg{appendix4} shows the resulting planets with this linear $L_{\star}$-$M_{\star}$ relation.  We find that the mass correlation between the planet and the star can still be obtained in Fig. \fgnum{appendix4}a and \fgnum{appendix4}b when two-component disks are considered. However, when the disks are purely stellar irradiated, the planet masses seem to be independent of their stellar hosts (Fig. \fgnum{appendix4}c and \fgnum{appendix4}d).  The latter independence can be derived from \eqs{h_irr}{rice_irr}, assuming a linear correlation between $L_{\star}$ and $M_{\star}$.    
 
 Planets around low-mass stars  in Fig. \fgnum{appendix4}a and \fgnum{appendix4}b  are drier compared to \fg{mass}. This is because these low-mass stars are brighter, and therefore the water ice lines in disks around such stars are in stellar irradiation regions.  Thus, $r_{\rm ice}$ moves outward with declining $\dot M_{\rm g}$ (see \eq{Psat} to \eq{P_eq}) while the planets mainly migrate inward. As a result, planets around such low-mass, bright stars accrete mostly dry pebbles and become rocky dominated.

\subsection{Neglecting the metallicity enhancement during the late stellar photoevaporation phase}
\label{ap:photoevaporation}
 
We emphasize that the disk metallicity is enhanced during the late  phase when disk gas is removed  through stellar photoevaporation. As a result, pebble accretion becomes enhanced, and the planet growth is boosted during this relatively short timespan.  Nevertheless, for a direct comparison, we also conduct simulations when disk effect is not taken into account, \ie , neglecting this metallicity enhancement in the stellar photoevaporation phase. The other parameters remain the same as in \tb{distribution}. The result is shown in  \fg{appendix5}.

Comparing \fg{appendix5} and Fig. \fgnum{mass}a and \fgnum{mass}b, we find that the overall mass scaling between planets and stars does not change. This process mostly affects the population of planets whose masses are lower than  $1 \Me$. Notice that there is a narrow concentration of sub-Earth mass planets with  low water fractions in Fig. \fgnum{mass}a and \fgnum{mass}b. 
This is because $r_{\rm ice}$ increases with declining $\dot M_{\rm g}$ in stellar photoevaporation phase. 
When including this process, the original low-mass, planets formed at the ice line undergo an additional  enhanced accretion of dry pebbles interior to the ice line,  and grow into slightly massive planets with low water fractions. 

\begin{figure}[t]
 \includegraphics[scale=0.7, angle=0]{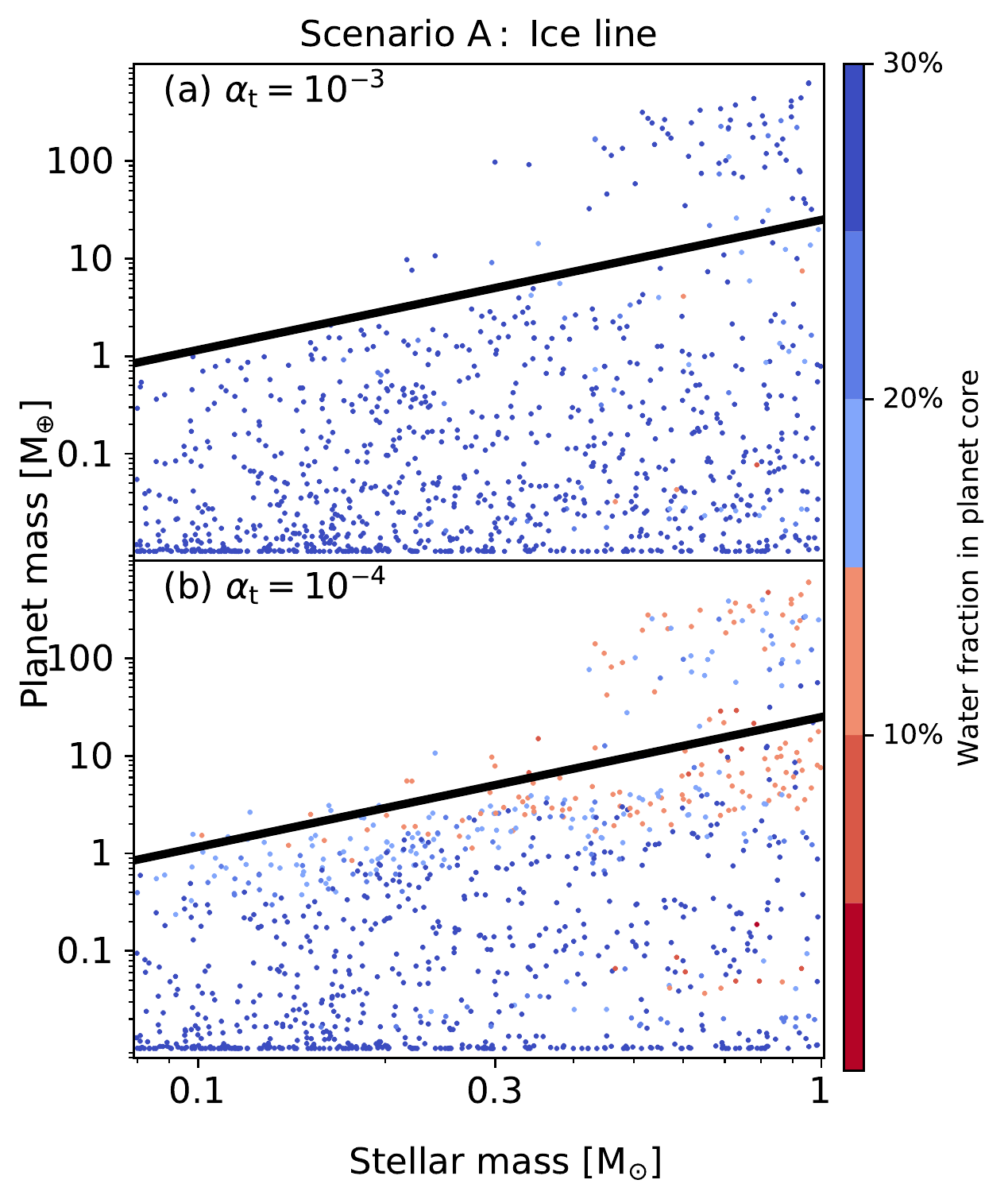}
       \caption{ 
       Same as Fig. \fgnum{mass}a and \fgnum{mass}b, but the disk metallicity is assumed to be the same during the stellar photoevaporation phase. Therefore, pebble accretion and the planet growth is not enhanced. But the masses of super Earth planets still linearly scale with their stellar masses.  
    }
\label{fig:appendix5}
\end{figure}

\subsection{Varying initial disk conditions}
\label{ap:disksize}
 
Due to the considerable uncertainties and scatterings among disk observations, we also explore another configuration of  initial disk conditions where disk sizes weakly correlate with their stellar masses $R_{\rm d0} \propto M_{\star}^{0.25}$ \citep{Bate2018}, and  $\dot M_{\rm g}\propto M_{\star}^{1.6}$ \citep{Hartmann1998,Alcala2014}. This also indicates a steeper than linear relation between disk masses and stellar masses.  The rest of the parameters remain the same as in \tb{distribution}. 

We find  in \fg{appendix6} that the planet distribution is quite similar to \fg{mass}.  The above linear mass correlation between planets and the stars can still be obtained. 

\begin{figure}[t]
 \includegraphics[scale=0.7, angle=0]{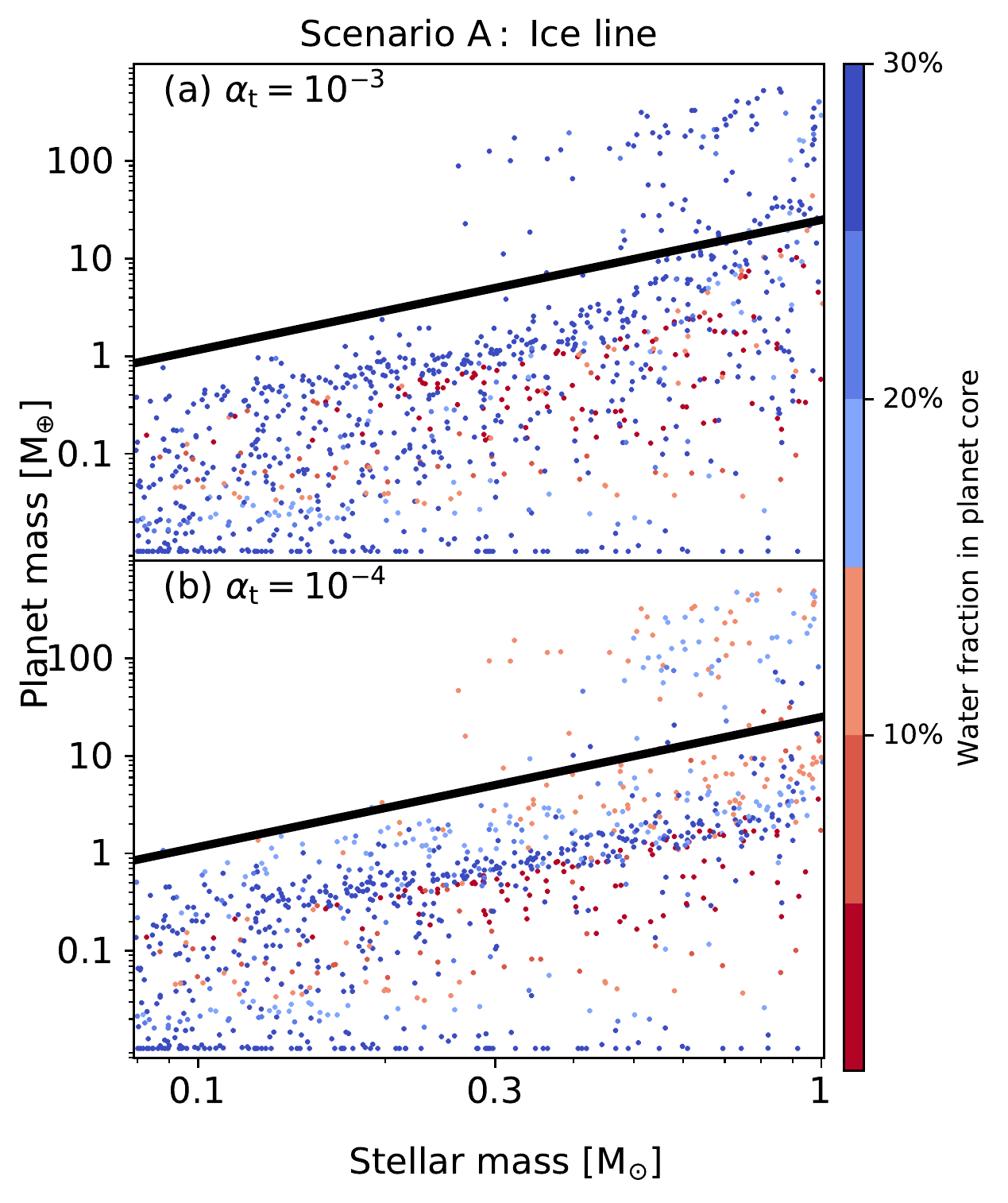}
       \caption{ 
       Same as Fig. \fgnum{mass}a and \fgnum{mass}b, but the initial disk conditions are assumed as $R_{\rm d0} \propto M_{\star}^{0.25}$ and $\dot M_{\rm g}\propto M_{\star}^{1.6}$. The masses of resulting planets also linearly scale with their stellar masses.  
    }
\label{fig:appendix6}
\end{figure}

\subsection{Low turbulent disk}
\label{ap:lowalpha}
 
It is plausible that protoplanetary disks have even lower $\alphat$ \citep{LiuSF2018,Zhang2018}. For instance, \cite{Zhang2018} reported a $0.1 \ M_{\rm J}$ planet in $\rm AS \ 209$ disk with $\alphat=10^{-5}$ actually best fit all the rings and gaps (their Fig. 19).  We also conduct simulations in this very low disk turbulence regime. The midplane gas velocity in calculating pebble accretion efficiency is also taken into account. Now the $2/3$D pebble accretion efficiencies are replaced by
 \begin{equation}
\begin{split}
\varepsilon_\mathrm{PA,2D} & =  \frac{0.32}{\eta +0.75 \alphag h_{\rm g}^2/\taus } \sqrt{ \frac{M_{\rm p} }{M_\star}  \frac{\Delta v}{ v_{\rm K} } \frac{1}{\taus}}  
\label{eq:eps-2D_new}
\end{split}
\end{equation}
and 
\begin{equation}
\label{eq:eps-3D}
\begin{split}
\varepsilon_\mathrm{PA, 3D} & =  \frac{0.39 }{(\eta +0.75 \alphag h_{\rm g}^2/\taus)} \frac{ 1}{ h_\mathrm{peb}} \frac{M_{\rm p}}{ M_\star}.
\end{split}
\end{equation}
 The rest of the parameters remain the same as in \tb{distribution}. 
 
\Fg{appendix7} shows the planet populations for both $\alphat=10^{-4}$ and $10^{-5}$. The disk turbulence has two effects. One the one hand, a lower $\alpha$ results in a smaller pebble scale height and a higher pebble accretion efficiency. The planet growth is faster.   On the other hand,  the outward migration effect is weaker for lower $\alpha$ disks. Planets mainly migrate inward and spend less time exterior to  the ice line in low turbulent disks, limiting the time for their mass growth.  Comparing these two effects, the overall planet distribution is quite similar between these two cases (\fg{appendix7}a and \fg{appendix7}b).  Super-Earths planets in disks of $\alpha=10^{-5}$ contain $5\%-10 \%$ water, drier than those formed in disks of $\alpha=10^{-4}$. There are fewer gas giant planets in \fg{appendix7}a compared to Fig. \fgnum{mass}b when the midplane gas velocity is included.  Importantly, the above linear mass correlation between planets and the stars is still obtained.

\begin{figure}[t]
 \includegraphics[scale=0.7, angle=0]{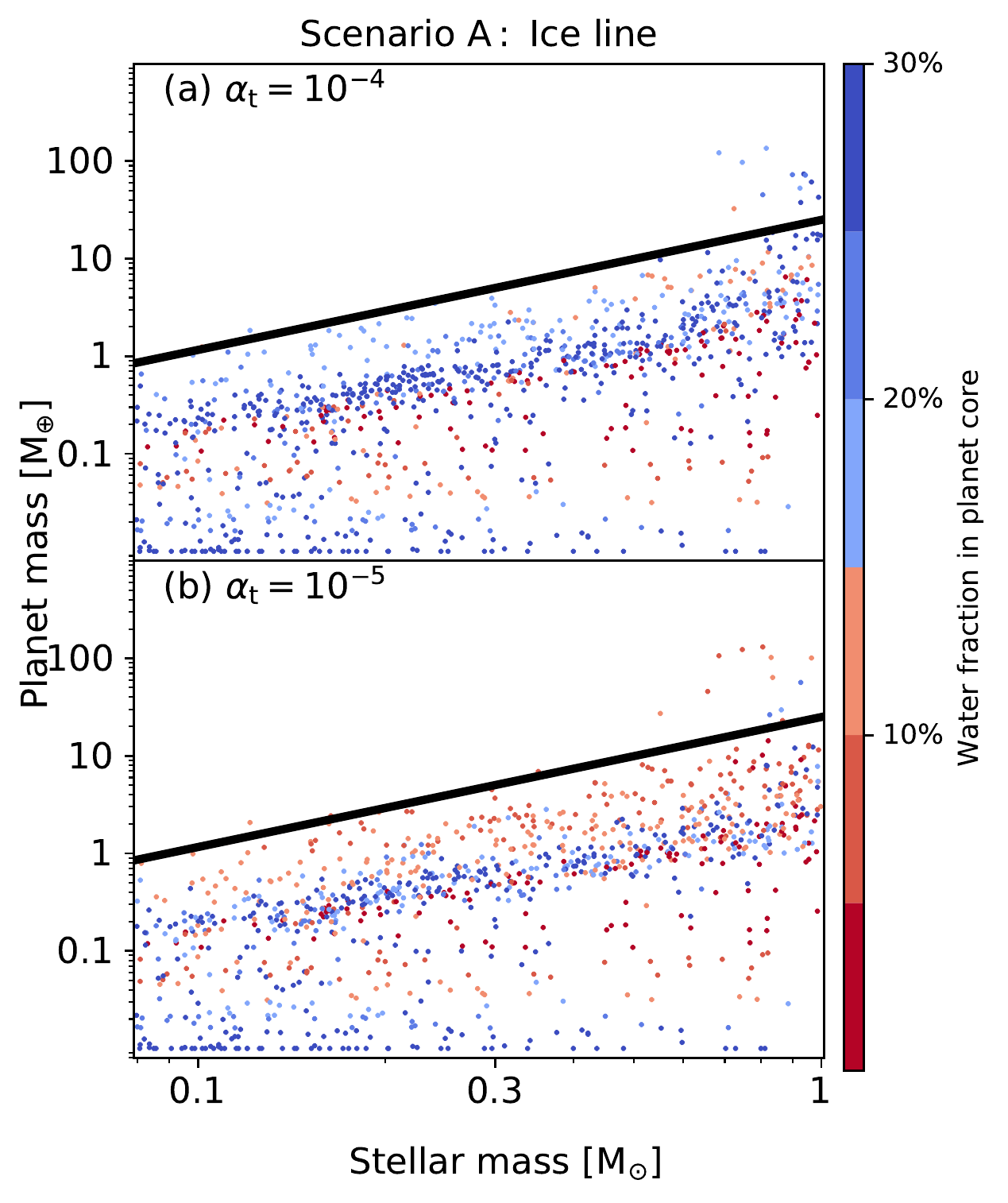}
       \caption{ 
    Same as Fig. \fgnum{mass}a and \fgnum{mass}b, but $\alphat$ is chosen to be $10^{-4}$ and $10^{-5}$. The midplane gas radial velocity is also included in calculating the pebble accretion efficiency.  Low turbulent disks generally produce planets with lower water fractions. Including the midplane gas velocity does not change the linear mass scaling between the planets and stars, but yields fewer gas giant planets.
     }
\label{fig:appendix7}
\end{figure}

\section{derivations for the disk structure of in the viscously heated region}
\label{ap:derivation}

In the inner viscously heated disk region, the energy balance between heating and cooling gives
\begin{align}
\frac{3}{4 \pi}  \dot{M_{\rm g}}   \Omegak^2 
= 2\sigma T_{\rm e}^{4},
\label{eq:energy}
\end{align}
where $T_{\rm e}$ is the effective disk temperature at the surface layer and the Stefan-Boltzmann constant $\sigma= 5.67\times 10^{-5} \rm \ g\, s^{-3} \, K^{-4}$. In an optically thick radiative disk,  the heat generated from viscous dissipation  occurs at the midplane, which radiates vertically towards  to the disk surface. The relation between the midplane temperature $T$ and the effective temperature $T_{\rm e}$ is given by   \cite{Hubeny1990}
\begin{equation}
T^{4} \simeq \frac{3}{8} \tau_{\rm  } T_{e}^{4}, 
\end{equation}
and the optical depth is
\begin{equation}
\tau_{\rm } = \kappa \Sigma_{\rm g}.
\end{equation}
Based on the above equations and  the assumed opacity law $\kappa =\kappa_{0}T$,
we obtain 
\begin{equation}
\frac{3}{4 \pi}  \dot{M_{\rm g}}   \Omegak^2 =  \frac{16 \sigma T ^{3} }{3 \kappa_{0}  \Sigma_{\rm g}}.
\label{eq:energy2}  
\end{equation}
The correlation between the disk aspect ratio $h_{\rm g}$, the gas surface density $\Sigma_{\rm g}$, and disk midplane temperature $T$ are given by  
\begin{equation}
H_{\rm g}^2=(h_{\rm g}r)^2=  c_{\rm s}^2/\Omega_{\rm K}^2= ( R_{\rm g}T/\mu)/ \Omega_{\rm K}^2,
  \label{eq:tempvis}
\end{equation}
and 
\begin{equation}
  \dot{M_{\rm g}} = 3\pi \nu \Sigma_{\rm g}= 3 \pi \alphag h_{\rm g}^2 r^2 \Omegak \Sigma_{\rm g},
   \label{eq:Sigmavis}
\end{equation} 
where  $R_{\rm g} = 8.31 \times 10^{7}  \ \rm J\, mol^{-1} \,K^{-1} $ is the gas constant and $\mu=2.33$ is the mean molecule weigh in the protoplanetary disk.
Inserted \eqs{tempvis}{Sigmavis} into \eq{energy2},   the disk aspect ratio  in the viscously heated region  can be expressed as 
\begin{equation}
h_{\rm g}= h_{\rm g, vis}=\left(   \frac{ 3  R_{\rm g}^3  \kappa_{0} \dot{M_{\rm g}}^2 }{ 64\pi^2 \sigma \mu^3   \alphag  \Omega_{\rm K}^{5} r^8 }     \right)^{1/8}
\end{equation} 

By given fiducial disk parameters, the above equations can be written as   
\begin{equation}
\begin{split}
h_{\rm g, vis} & \simeq 3.4 \times 10^{-2}  \left(\frac{M_{\star}}{ M_{\odot}} \right)^{-5/16}  \left( \frac{\dot{M} }{ 10^{-8}  
 \rm M_{ \odot}\, yr^{-1}  } \right)^{1/4} \\
 &   \left(\frac{\alphag}{10^{-2}}\right)^{-1/8}  \left(\frac{ r}{1 \rm AU}  \right)^{-1/16}.  
 \label{eq:hvis}
\end{split}
\end{equation}
The surface density and mid-plane temperature  are calculated accordingly by replacing \eq{hvis} into \eqs{Sigmavis}{tempvis}, 
\begin{equation}
\begin{split}
    \Sigma_{\rm g,vis} = &  132
    \left( \frac{\dot M_{\rm g}}{10^{-8} \Msyr} \right)^{1/2}  \left(\frac{M_{\star}}{1 \ M_{\odot}} \right)^{1/8}   
    \left(\frac{\alphag} {10^{-2}} \right)^{-3/4}\\
    & \left(\frac{\kappa_0}{10^{-2}} \right)^{-1/4} 
    \left(\frac{r}{1 \AU} \right)^{-3/8}
    \ \rm gcm^{-2},
    \end{split}
    \end{equation}
    and 
\begin{equation}
\begin{split}    
T_{\rm g,vis} = & 280
\left( \frac{\dot M_{\rm g}}{10^{-8} \Msyr} \right)^{1/2}
\left(\frac{M_{\star}}{1 \ M_{\odot}} \right)^{3/8} 
\left(\frac{\alphag} {10^{-2}} \right)^{-1/4}\\
& \left(\frac{\kappa_0}{10^{-2}} \right)^{1/4} 
\left(\frac{r}{1 \AU} \right)^{-9/8}
\ \rm K.
  \end{split}
\end{equation}

\begin{acknowledgements}
We thank Chris Ormel, Gijs Mulders, Thomas Ronnet and Bertram Bitsch for useful discussions. We also thank the anonymous referee for their useful suggestions and comments. 
B.L. is supported by the European Research Council (ERC Consolidator Grant 724687-PLANETESYS) and the Swedish Walter Gyllenberg Foundation. M.L. is funded by the Knut and Alice Wallenberg Foundation (Wallenberg Academy Fellow Grant 2012.0150). A.J. thanks the European Research Council (ERC Consolidator Grant 724687-PLANETESYS), the Knut and Alice Wallenberg Foundation (Wallenberg Academy Fellow Grant 2012.0150) and the Swedish Research Council (Project Grant 2018-04867) for research support.
F. L. is supported by the grant "The New Milky Way" from the Knut and Alice Wallenberg Foundation and the grant 184/14 from the Swedish National Space Agency.
The computations were performed on resources provided by the Swedish Infrastructure for Computing (SNIC) at the LUNARC-Centre in Lund.
\end{acknowledgements}

\bibliographystyle{aa}
\bibliography{reference}

\end{document}